\documentclass[traditabstract]{aa} % for the abstract without structuration 
\usepackage{graphicx,natbib}
\usepackage{txfonts}
\bibpunct{(}{)}{;}{a}{}{,}
\begin{document}
   \title{Mn, Cu, and Zn abundances in barium stars and their correlations
          with neutron capture elements\thanks{Based on spectroscopic
          observations collected at the European Southern Observatory (ESO),
          within the Observat\'orio Nacional ON/ESO and ON/IAG agreements,under
          FAPESP project n$^{\circ}$ 1998/10138-8.}\fnmsep\thanks{Tables \ref{abun}
          and \ref{medxfe}, are only available in electronic form at the CDS via
          anonymous ftp to cdsarc.u-strasbg.fr (130.79.128.5) or via
          http://cdsweb.u-strasbg.fr/cgi-bin/qcat?J/A+A/}\fnmsep\thanks{
	  Figures \ref{12356fg1}, \ref{12356fg2}, \ref{12356fg3}, 
	  \ref{12356fg4}, \ref{12356fg5}, \ref{12356fg7}, and \ref{12356fg8}
	  and Tables \ref{patm}, \ref{lit}, and \ref{fits} are only available in 
	  electronic form at http://www.edpsciences.org}}

%   \subtitle{}

   \author{D. M. Allen
          \inst{1}
      \and G. F. Porto de Mello\inst{2}
          }

   \institute{Instituto de Astronomia, Geof\'\i sica e Ci\^encias
              Atmosf\'ericas, Universidade de S\~ao Paulo, Rua do Mat\~ao
              1226, CEP: 05508-900, S\~ao Paulo, Brazil \\
          \email{dimallen@astro.iag.usp.br}
      \and
          Universidade Federal do Rio de Janeiro, Observat\'orio do Valongo,
          Ladeira do Pedro Antonio 43, CEP: 20080-090, Rio de Janeiro, RJ, Brazil \\
          \email{gustavo@astro.ufrj.br}
             }

   \date{Received; accepted}
 
\authorrunning{Allen \& Porto de Mello}

\titlerunning{Mn, Cu, Zn: correlations with neutron capture elements}

% \abstract{}{}{}{}{} 
% 5 {} token are mandatory

  \abstract
{Barium stars are optimal sites for studying the correlations
between the neutron-capture elements and other species that may
be depleted or enhanced, because they act as neutron seeds or poisons 
during the operation of the $s$-process. These data are necessary to
help constrain the modeling of the neutron-capture paths and
explain the $s$-process abundance curve of the solar system. Chemical
abundances for a large number of barium stars with different
degrees of $s$-process excesses, masses, metallicities, and
evolutionary states are a crucial step towards this goal. We
present abundances of Mn, Cu, Zn, and various light and
heavy elements for a sample of barium and normal giant stars, and
present correlations between abundances contributed to
different degrees by the weak-$s$, main-$s$, and $r$-processes of
neutron capture, between Fe-peak elements and heavy elements.
Data from the literature are also considered in order to better
study the abundance pattern of peculiar stars. The stellar spectra
were observed with FEROS/ESO. The stellar atmospheric
parameters of the eight barium giant stars and six normal giants
that we analyzed lie in the range 4300 $<$ $T_{\rm eff}$/K $<$ 5300,
$-$0.7 $<$ [Fe/H] $\leq$ 0.12 and 1.5 $\leq \log g <$ 2.9. Carbon
and nitrogen abundances were derived by spectral synthesis of the
molecular bands of C$_2$, CH, and CN. For all other elements
we used the atomic lines to perform the spectral synthesis. A very
large scatter was found mainly for the Mn abundances
when data from the literature were considered. We found that [Zn/Fe]
correlates well with the heavy element excesses, its abundance
clearly increasing as the heavy element excesses increase, a trend
not shown by the [Cu/Fe] and [Mn/Fe] ratios. 
Also, the ratios involving Mn, Cu, and Zn and heavy elements 
usually show an increasing trend toward higher metallicities.
Our results suggest
that a larger fraction of the Zn synthesis than of Cu is owed to massive
stars, and that the contribution of the
main-s process to the synthesis of both elements is small. We also
conclude that Mn is mostly synthesized by SN Ia, and that a
non-negligible fraction of the synthesis of Mn, Cu, and Zn is owed
to the weak s-process.}


   \maketitle
%
%________________________________________________________________

\section{Introduction}\label{int}

It is becoming increasingly clear in the present age of
high-resolution spectroscopy of large databases and precise
abundances for numerous chemical elements that the galactic
chemical evolution (GCE) models are lacking with regard to the
available observational data. A successful GCE model must
incorporate stellar evolution inputs such as an initial mass
function, star-formation rate, mass loss through winds and
eruptive processes, and also differing timescales for stellar
nucleosynthetic yields and their sensitivities to differing
metallicities. It must also link these to the complexities of
dynamical galactic components and the great diversity of observed
structures, in which formation, dissolution, and merging processes
labor many hierarchical levels. Such a model is largely
lacking, and it must involve a sufficiently large number of still
essentially constraint-free parameters that the most stringent
challenges presently rest on the theoretical side, instead of the
observational one.

A good example of this state of affairs is the nucleosynthesis of
Mn, Cu, and Zn. 
The astrophysical sites of, timescales of, and
dominant contributing processes to the synthesis of the Fe-peak
element Mn, a classical Fe-peak element, and Cu and Zn, considered
to be the lightest $s$-process main component nuclei and the heaviest
Fe-peak element respectively, still have not been reliably established.
A theoretical and observational debate is still on-going to determine the
expected main contributors to the observed Mn abundances, whether
type II or type Ia supernovae (SNe), and to establish if the Mn
yields depend on the metallicity of the SN progenitors
\citep{feltzing07,nissen00}.

Cu and Zn are part of an even more complex scenario. They have
traditionally been considered as transition elements between the
Fe-peak and the light $s$-process species \citep{sneden91}. They are
thought to be produced through a variety of nucleosynthetic
processes, including explosive nucleosynthesis in SN II and SN Ia,
the main component of the $s$-process, thought to occur in the
He-burning shells of low to intermediate mass AGB stars, and the
weak component of the $s$-process, thought to be sited at
He-burning cores of M $\geq$ 10 M$_\odot$ stars \citep[see][and
references therein]{raiteri93,matte93,mish02}. Even though their 
positions in the periodic table are contiguous, their behavior 
in the GCE scenario is in sharp contrast.

The solar curves for the abundance of $s$-only nuclei vs. mass number
of \citet{kapp89} show that the observational data very well fit
the theoretical curve for nuclei from A $>$ 90, whereas for
lighter nuclei another process is necessary to explain the solar
abundances. This process was called the weak component of the
$s$-process, whose reproduction by successive stellar generations 
was investigated
taking into account the interference of the main component in the
atomic mass range from Fe to Sr, and a satisfactory match to the
solar $s$-abundance was reached. \citet{lamb77} studied the
contribution of the weak component for nuclei lighter than Fe and
found that significant amounts of some elements with A $<$ 60 are
produced in stars with M $<$ 15M$_\odot$ and that nuclei in the atomic
range 60 $\leq$ A $\leq$ 70 are made in stars more massive than
about 15M$_\odot$.

Given the very intricate scenario for the nucleosynthesis of Mn,
Cu, and Zn, additional abundance data from as many nucleosynthetic
sources as possible are expected to contribute to the
quantification of the various processes at play. Our aim in
this paper is to help clarify the nucleosynthetic processes that 
contribute to the production of Mn, Cu, and Zn by comparing their
abundances in a sample of barium dwarf and giant stars to
the abundances of heavy elements. Barium stars show
high excesses of the elements produced by the main $s$-process.
Because they are not evolved enough to have self-enriched in
these elements, a binary nature is invoked to explain their
existence \citep{mcclureetal1980}. Thus, chemical peculiarities of
the barium stars come from a more massive companion, presently
observed (if at all) as a white dwarf. It evolved faster and, as a
thermally pulsating AGB, became self-enriched with products
of the main $s$-process, thereby transferring this enriched
material onto the current barium star after dredge-up episodes
through a stellar wind. These overabundances, then, are not
intrinsic to the barium star, can be directly linked to the
AGB nucleosynthesis, and therefore are important tests of
the theories of nucleosynthesis and the chemical evolution of the
Galaxy. The main goal of our effort is to provide new
observational constraints to the chemical evolution of Mn, Cu, and
Zn by gauging the effects on these elements of $s$-process
nucleosynthesis in heavy-element enhanced stars.

The paper is organized as follows: Section \ref{param} briefly presents
the data and the determination of the stellar atmospheric
parameters; Sect. \ref{abunt} briefly describes the abundance and
uncertainty determinations; in Sect. \ref{disc} the correlations
between the Mn, Cu, and Zn abundances and those of the neutron
capture elements are discussed, and in Sect. \ref{concl}
we draw our conclusions.

%__________________________________________________________________

\section{Observations and atmospheric parameters}\label{param}

All spectra for the sample stars were obtained with the 1.52m
telescope at ESO, La Silla, using the Fiber Fed Extended Range
Optical Spectrograph \citep[FEROS,][]{kaufer00}. FEROS spectra
have a constant resolving power of R = 48000 from 3600 $\rm \AA$
to 9200 $\rm \AA$. The stellar sample targeted in our 
study includes eight mild and classical barium stars and six
normal giants of \citet{rod07}, with a spectral S/N ratio ranging
from 200 to 450 in the visible range.

\citet{rod07} determined $T_{\rm eff}$, $\log g$, and the metallicities
from the simultaneous excitation and ionization equilibria of the
equivalent widths of an average number of 120 \ion{Fe}{i} and 12
\ion{Fe}{ii} lines. Surface gravities were also computed from the
stellar luminosities and theoretical HR diagrams, and a very good
agreement was found for the two sets of gravities. A summary of
the stellar parameters is given in Table \ref{patm}. The reader is
referred to \citet{rod07} for a journal of the observations,
details of derivation of atmospheric parameters, and the respective
uncertainties.

Additionally, we determined the Mn abundances for the 26 Ba
stars analyzed by \citet{di06a}, to whom we refer the reader for
details on observations and the derivation of atmospheric
parameters.

%__________________________________________________________________

%------------Table param------------------------------------------------------------
\onltab{1}{
\begin{table*}
\caption{Stellar parameters for the sample stars derived by
\citet{rod07}. T$_{exc}$: excitation temperature; $\log g$:
surface gravity; [Fe/H]: metallicities; $\xi$: microturbulent
velocities; BC(V): bolometric correction; M$_v$: absolute
magnitudes; M$_{bol}$: bolometric magnitude; $L_\ast/L_\odot$:
luminosities; $M_\ast/M_\odot$: masses. Numbers in parenthesis are
errors in last decimals.} 
\label{patm}
\centering
\begin{tabular}{lccllllllc}
\hline\hline
Star & T$\rm _{exc}$ & $\log g$ & [Fe/H] & $\xi$ & BC(V) & M$_v$ & M$_{bol}$ & $L_\ast/L_\odot$ & $M_\ast/M_\odot$ \\
& (K) & (cgs) & & (km/s) & &&&& \\
\hline
HD 9362    & 4780(50) & 2.43(35) & $-$0.34(7)  & 1.71(6) & $-$0.34 &   +0.78 &   +0.44(4)  & 1.75(2)  & 1.9 \\
HD 13611   & 5120(50) & 2.49(35) & $-$0.14(6)  & 1.96(6) & $-$0.22 & $-$0.76 & $-$0.98(21) & 2.35(9)  & 3.6 \\
HD 20894   & 5080(50) & 2.60(35) & $-$0.11(6)  & 1.80(6) & $-$0.23 &   +0.04 & $-$0.19(14) & 2.00(5)  & 3.0 \\
HD 26967   & 4650(50) & 2.51(35) &    0.00(7)  & 1.52(6) & $-$0.39 &   +1.21 & $-$0.81(3)  & 1.60(1)  & 1.5 \\
HD 46407   & 4940(50) & 2.65(35) & $-$0.09(12) & 1.87(6) & $-$0.28 &   +0.97 &   +0.68(16) & 1.65(7)  & 2.3 \\
HD 104979  & 4920(50) & 2.58(35) & $-$0.35(5)  & 1.71(6) & $-$0.29 &   +0.63 &   +0.34(6)  & 1.79(3)  & 2.3 \\
HD 113226  & 5082(50) & 2.85(35) &   +0.12     & 1.86(6) & $-$0.23 &   +0.47 &   +0.24(4)  & 1.83(2)  & 2.9 \\
HD 116713  & 4790(50) & 2.67(35) & $-$0.12(13) & 1.97(6) & $-$0.34 &   +1.23 &   +0.89(8)  & 1.57(3)  & 1.9 \\
HD 139195  & 5010(50) & 2.89(35) & $-$0.02(6)  & 1.67(6) & $-$0.26 &   +1.07 &   +0.82(9)  & 1.60(4)  & 2.4 \\
HD 181053  & 4810(50) & 2.48(35) & $-$0.19(6)  & 1.70(6) & $-$0.33 &   +0.64 &   +0.31(21) & 1.80(8)  & 2.2 \\
HD 202109  & 4910(50) & 2.41(35) & $-$0.04(7)  & 1.85(6) & $-$0.29 & $-$0.01 & $-$0.31(3)  & 2.05(1)  & 3.0 \\
HD 204075  & 5250(50) & 1.53(35) & $-$0.09(12) & 2.49(6) & $-$0.18 & $-$1.57 & $-$1.76(18) & 2.63(7)  & 4.2 \\
HD 205011  & 4780(50) & 2.41(35) & $-$0.14(9)  & 1.70(6) & $-$0.34 &   +0.54 &   +0.20(15) & 1.85(6)  & 2.2 \\
HD 220009  & 4370(50) & 1.91(35) & $-$0.67(7)  & 1.61(6) & $-$0.52 &   +0.11 & $-$0.41(13) & 2.09(5)  & 1.0 \\
\noalign{\smallskip}
\hline
\end{tabular}
\end{table*}
}
%-------------------------------------------------------------------------------

%-------Table abun (CDS)------------------------------------------------------------------------
\begin{table}
\caption{Lines, equivalent widths, and abundances.
Abundances of Mn for barium stars from \citet{di06a} are shown
starting on line 1541. Note that HD 2454 is HR 107 in
\citet{di06a}. The full table is only available in CDS.}
{\scriptsize 
\label{abun}
\centering
\setlength\tabcolsep{3pt}
\begin{tabular}{lllcccccc}
\hline\hline
Star & El & $\lambda$ & $\chi\sb {ex}$ & log gf & Ref & EW & $\log\epsilon$ & [X/Fe] \\
&& ($\rm \AA$) & (eV) & & & (m$\rm\AA$) &  & \\
\hline
HD 9362* & Li I **& 6707.760  & 0.000 &  0.171 & 1  & ... & -0.34 & ... \\
HD 9362* & Li I **& 6707.910  & 0.000 & -0.299 & 2  & ... & -0.34 & ... \\
HD 9362* & Li I **& 6708.087  & 0.000 &  0.000 & 3  & ... & -0.34 & ... \\
HD 9362* & C      & 4295.000a &  ...  & ...    & ...& ... &  8.16 & -0.02 \\
HD 9362* & C      & 5135.600b &  ...  & ...    & ...& ... &  8.19 &  0.01 \\
\noalign{\smallskip}
\hline
\end{tabular}
}
\tablefoot{
``*'' indicates stars that were considered to be normal 
instead of barium stars by \citet{rod07}, and the well-known giant
$\epsilon$ Vir (HD 113226), used by them as reference star; ``**'' indicates multiple line;
``a'': representative line for CH (G band); ``b'': representative line for C$_2$;
``c'': representative line for CN.
$<$ indicates an upper limit. The gf-values sources are given below.
}
\tablebib{
(1) \citet{M94}; (2) NIST; (3) \citet{barb99}; (4) \citet{allende01};
(5) \citet{L78}; (6) \citet{Bi75}; (7) \citet{BG80}; (8) \citet{g94};
(9) \citet{H82}; (10) \citet{HL83}; (11) \citet{B81}; (12) \citet{T90};
(13) \citet{T89}; (14) \citet{S00}; (15) VALD; (16) \citet{M98};
(17) \citet{rut78}; (18) \citet{law01a}; (19) \citet{PQ00}; (20) \citet{G91};
(21) \citet{L76}; (22) \citet{Hartog03}; (23) \citet{MW77}; (24) \citet{S96};
(25) \citet{B89}; (26) \citet{law01b}; (27) \citet{B88};
(28) \citet{C62}; (29) average between \citet{K92} and \citet{BL93};
(30) \citet{Biemont00}; ``SUN'' indicates $\log gf$ obtained through fits
on the solar spectrum.
}
\end{table}
%-------------------------------------------------------------------------------

%------------Table medxfe (CDS)--------------------------------------------------------------------------------------------------------
\begin{table}
\caption{Average for $\log\epsilon$(X) and [X/Fe]. Uncertainties on
$\log\epsilon$(X) and [X/Fe] are represented by $\sigma_l$, and
$\sigma_f$, respectively. Given that \citet{rod07} used HD 113226
as a reference star, there is no uncertainty designed to it.
The full table is only available in CDS.
} 
\label{medxfe}
\setlength\tabcolsep{3pt}
\begin{tabular}{llrcrc}
\hline\hline
Star & El & $\log\epsilon$(X) & [X/Fe] & $\sigma_l$ & $\sigma_f$ \\
\hline
HD 9362* & Li & -0.34 & ...  & 0.14 & 0.16 \\
HD 9362* &  C &  8.19 & 0.01 & 0.10 & 0.10 \\
HD 9362* &  N &  7.73 & 0.15 & 0.09 & 0.08 \\
HD 9362* &  O &  8.54 & 0.14 & 0.17 & 0.17 \\
HD 9362* & Mn &  4.80 &-0.25 & 0.11 & 0.10 \\
HD 9362* & Cu &  3.66 &-0.21 & 0.12 & 0.11 \\
HD 9362* & Zn &  4.01 &-0.25 & 0.13 & 0.13 \\
\noalign{\smallskip}
\hline
\end{tabular}
\tablefoot{
The symbol ``*'' indicates those stars that were considered to be normal 
instead of barium stars by \citet{rod07}. 
From the line 351, the data of
\citet{di06a} with corrected uncertainties are shown. 
From the line 1001, the solar abundances and their references are shown.
}
\end{table}
%--------------------------------------------------------------------------------------------------------------------------------

%------------Table hfsMnCu------------------------------------------------------------
\begin{table}
\caption{Hyperfine structure for \ion{Mn}{i} and \ion{Cu}{i} lines.} 
\label{hfsMnCu} 
\centering
\setlength\tabcolsep{2pt}
\begin{tabular}{ccccccc}
\noalign{\smallskip}
\hline\hline
ID & $\lambda$ ($\rm \AA$) & $\log gf$ && ID & $\lambda$ ($\rm \AA$) & $\log gf$ \\
\noalign{\smallskip}
\cline{1-3} \cline{5-7} \\
\multicolumn{3}{c}{4754.042$\rm \AA$;  $\chi_{ex}$=2.282 eV} && \multicolumn{3}{c}{6021.792$\rm \AA$; $\chi_{ex}$=3.075 eV} \\
\multicolumn{3}{c}{$\log gf$(total) = -0.086\tablefootmark{a}} && \multicolumn{3}{c}{$\log gf$(total) = -0.216\tablefootmark{a}} \\
\noalign{\smallskip}
\cline{1-3} \cline{5-7} \\
\ion{Mn}{i} & 4754.0210 & -0.6523 &&  \ion{Mn}{i} & 6021.7190 & -2.6674 \\
\ion{Mn}{i} & 4754.0320 & -0.7918 &&  \ion{Mn}{i} & 6021.7450 & -1.4499 \\
\ion{Mn}{i} & 4754.0420 & -0.9502 &&  \ion{Mn}{i} & 6021.7480 & -2.3152 \\
\ion{Mn}{i} & 4754.0520 & -1.1342 &&  \ion{Mn}{i} & 6021.7680 & -1.2739 \\
\ion{Mn}{i} & 4754.0540 & -1.5699 &&  \ion{Mn}{i} & 6021.7710 & -2.1903 \\
\ion{Mn}{i} & 4754.0600 & -1.3561 &&  \ion{Mn}{i} & 6021.7760 & -0.5323 \\
\ion{Mn}{i} & 4754.0600 & -1.3939 &&  \ion{Mn}{i} & 6021.7870 & -1.2483 \\
\ion{Mn}{i} & 4754.0640 & -1.3683 &&  \ion{Mn}{i} & 6021.7910 & -2.2695 \\
\ion{Mn}{i} & 4754.0660 & -1.6413 &&  \ion{Mn}{i} & 6021.7940 & -0.6718 \\
\ion{Mn}{i} & 4754.0680 & -1.4352 &&  \ion{Mn}{i} & 6021.8000 & -1.3152 \\
\ion{Mn}{i} & 4754.0710 & -1.6113 &&  \ion{Mn}{i} & 6021.8070 & -0.8302 \\
\ion{Mn}{i} & 4754.0800 & -2.3895 &&  \ion{Mn}{i} & 6021.8100 & -1.4913 \\
\ion{Mn}{i} & 4754.0820 & -2.3103 &&  \ion{Mn}{i} & 6021.8160 & -1.0142 \\
\ion{Mn}{i} & 4754.0820 & -2.4352 &&  \ion{Mn}{i} & 6021.8200 & -1.2361 \\
\ion{Mn}{i} & 4754.0820 & -2.7874 &&  \ion{Mn}{i} & 6021.8200 & -1.5213 \\
\multicolumn{3}{c}{\hbox to 4.0cm {\hrulefill}} && \multicolumn{3}{c}{\hbox to 4.0cm {\hrulefill}} \\
\multicolumn{3}{c}{5420.350$\rm \AA$;  $\chi_{ex}$=2.143 eV} && \multicolumn{3}{c}{5105.50$\rm \AA$;  $\chi_{ex}$=1.39 eV} \\
\multicolumn{3}{c}{$\log gf$(total) = -1.460\tablefootmark{a}} && \multicolumn{3}{c}{$\log gf$(total) = -1.520\tablefootmark{b}} \\
\ion{Mn}{i} & 5420.2600 & -3.0163 &&  \ion{Cu}{i} & 5105.497 & -4.2291  \\
\ion{Mn}{i} & 5420.2650 & -2.9863 &&  \ion{Cu}{i} & 5105.501 & -3.2774  \\
\ion{Mn}{i} & 5420.2730 & -2.7311 &&  \ion{Cu}{i} & 5105.503 & -3.2314  \\
\ion{Mn}{i} & 5420.2730 & -3.7645 &&  \ion{Cu}{i} & 5105.504 & -4.4202  \\
\ion{Mn}{i} & 5420.2820 & -2.8102 &&  \ion{Cu}{i} & 5105.510 & -3.1656  \\
\ion{Mn}{i} & 5420.2960 & -2.5092 &&  \ion{Cu}{i} & 5105.514 & -2.9097  \\
\ion{Mn}{i} & 5420.2960 & -3.6853 &&  \ion{Cu}{i} & 5105.519 & -3.8761  \\
\ion{Mn}{i} & 5420.3090 & -2.7433 &&  \ion{Cu}{i} & 5105.523 & -2.9318  \\
\ion{Mn}{i} & 5420.3270 & -2.3252 &&  \ion{Cu}{i} & 5105.525 & -2.7319  \\
\ion{Mn}{i} & 5420.3280 & -3.8102 &&  \ion{Cu}{i} & 5105.526 & -4.0655 \\
\ion{Mn}{i} & 5420.3450 & -2.7689 &&  \ion{Cu}{i} & 5105.530 & -2.6600 \\
\ion{Mn}{i} & 5420.3680 & -2.1668 &&  \ion{Cu}{i} & 5105.531 & -2.8199 \\
\ion{Mn}{i} & 5420.3690 & -4.1624 &&  \ion{Cu}{i} & 5105.534 & -2.5629 \\
\ion{Mn}{i} & 5420.3910 & -2.9449 &&  \ion{Cu}{i} & 5105.545 & -2.9140 \\
\ion{Mn}{i} & 5420.4180 & -2.0273 &&  \ion{Cu}{i} & 5105.550 & -2.3138 \\
\multicolumn{3}{c}{\hbox to 4.0cm {\hrulefill}} &&  \ion{Cu}{i} & 5105.554 & -2.4538 \\
\multicolumn{3}{c}{5432.548 $\rm \AA$;  $\chi_{ex}$=0.000 eV} &&  \ion{Cu}{i} & 5105.572 & -2.1079 \\
\multicolumn{3}{c}{$\log gf$(total) = -3.795\tablefootmark{a}} && \multicolumn{3}{c}{\hbox to 4.0cm {\hrulefill}} \\
\ion{Mn}{i} & 5432.5080 & -4.3769 &&  \multicolumn{3}{c}{5218.21$\rm \AA$; $\chi_{ex}$=3.82 eV} \\
\ion{Mn}{i} & 5432.5120 & -5.1550 &&  \multicolumn{3}{c}{$\log gf$(total) = +0.264\tablefootmark{a}} \\
\ion{Mn}{i} & 5432.5370 & -5.1550 &&   \ion{Cu}{i} & 5218.201 & -1.4123 \\
\ion{Mn}{i} & 5432.5400 & -4.6401 &&   \ion{Cu}{i} & 5218.203 & -0.9367 \\
\ion{Mn}{i} & 5432.5430 & -4.9923 &&   \ion{Cu}{i} & 5218.205 & -1.0665 \\
\ion{Mn}{i} & 5432.5620 & -4.9923 &&   \ion{Cu}{i} & 5218.207 & -0.3466 \\
\ion{Mn}{i} & 5432.5650 & -4.9711 &&   \ion{Cu}{i} & 5218.211 & -0.5575 \\
\ion{Mn}{i} & 5432.5670 & -4.9869 &&   \ion{Cu}{i} & 5218.213 & -0.5685 \\
\ion{Mn}{i} & 5432.5820 & -4.9869 &&   \ion{Cu}{i} & 5218.216 & -0.2226 \\
\ion{Mn}{i} & 5432.5840 & -5.4182 && \multicolumn{3}{c}{\hbox to 4.0cm {\hrulefill}} \\
\ion{Mn}{i} & 5432.5860 & -5.0892 &&  \multicolumn{3}{c}{5782.14 $\rm \AA$;  $\chi_{ex}$=1.64 eV} \\
\ion{Mn}{i} & 5432.5950 & -5.0892 &&  \multicolumn{3}{c}{$\log gf$(total) = -1.720\tablefootmark{b}} \\
\ion{Mn}{i} & 5432.5970 & -6.1172 &&   \ion{Cu}{i} & 5782.032 & -3.4320 \\
\ion{Mn}{i} & 5432.5980 & -5.3513 &&   \ion{Cu}{i} & 5782.042 & -3.7349 \\
\ion{Mn}{i} & 5432.6030 & -5.3513 &&   \ion{Cu}{i} & 5782.054 & -3.0340 \\
\multicolumn{3}{c}{\hbox to 4.0cm {\hrulefill}}                       &&   \ion{Cu}{i} & 5782.064 & -3.0850 \\
\multicolumn{3}{c}{6013.488$\rm \AA$;  $\chi_{ex}$=3.072 eV} && \ion{Cu}{i} & 5782.073 & -3.3890 \\
\multicolumn{3}{c}{$\log gf$(total) = -0.252\tablefootmark{a}}            && \ion{Cu}{i} & 5782.084 & -2.6880 \\
\ion{Mn}{i} & 6013.4530 & -0.7669 && \ion{Cu}{i} & 5782.086 & -3.0340 \\
\ion{Mn}{i} & 6013.4740 & -0.9790 && \ion{Cu}{i} & 5782.098 & -3.0340 \\
\ion{Mn}{i} & 6013.4920 & -1.2520 && \ion{Cu}{i} & 5782.113 & -2.6880 \\
\ion{Mn}{i} & 6013.5010 & -1.4561 && \ion{Cu}{i} & 5782.124 & -2.6880 \\
\ion{Mn}{i} & 6013.5080 & -1.6622 && \ion{Cu}{i} & 5782.153 & -2.5870 \\
\ion{Mn}{i} & 6013.5130 & -1.3100 && \ion{Cu}{i} & 5782.173 & -2.2410 \\
\ion{Mn}{i} & 6013.5210 & -1.3312 &&                      \\
\ion{Mn}{i} & 6013.5270 & -1.4861 &&                      \\
\ion{Mn}{i} & 6013.5370 & -1.8083 &&                      \\
\ion{Mn}{i} & 6013.5410 & -1.8541 &&                      \\
\ion{Mn}{i} & 6013.5410 & -2.4104 &&                      \\
\ion{Mn}{i} & 6013.5420 & -2.0302 &&                      \\
\noalign{\smallskip}
\hline
\end{tabular}
\tablefoot{$\log gf$ total sources: 
\tablefoottext{a}{NIST};
\tablefoottext{b}{\citet{Bi75}}
}
\end{table}
%-------------------------------------------------------------------------------

%__________________________________________________________________

\section{Abundances}\label{abunt}

In the present work we derived the abundances based on a spectrum
synthesis for 14 stars that were previously analyzed by \citet{rod07}.
Although the focus of this paper is on performing relations between
abundances of some iron peak and heavy elements, other elements 
than those discussed here are also shown in tables, 
in order to assemble all abundances determined so far for
these stars in a single source, as explained in Sects. \ref{ablight}
and \ref{abheavy}. Smiljanic and co-authors had already
derived abundances for some of these elements through equivalent
widths, but more accurate results are expected from spectrum
synthesis both for heavy elements with weak and very strong
lines. The abundance results line by line are shown in Table
\ref{abun}, and the average abundances are shown in Table
\ref{medxfe}. Abundances of Mn that were previously unpublished 
as well as new calculations for the abundance uncertainties  
for the 26 stars of \citet{di06a} are also
shown in Tables \ref{abun} and \ref{medxfe}. Details of the
sample, the atmospheric parameters, and the abundances derivation are
described there.

\citet{rod07} and \citet{di06a} employ different methods to
obtain the stellar atmospheric parameters T$_{\rm eff}$, $\log g$,
[Fe/H] and microturbulence, so we needed to establish a 
satisfactory consistence between the two sets of parameters, 
essentially the T$_{\rm eff}$ scales. \citet{rod07} rely on 
the excitation and ionization equilibria 
of \ion{Fe}{i}/\ion{Fe}{ii} to derive the
parameters self-consistently from the spectroscopic data alone,
while \citet{di06a} derive T$_{\rm eff}$ from photometry and the
surface gravity from the stellar positions in the HR diagram.
\cite{di06a}, however, also obtain excitation and ionization
T$_{\rm eff}$ values, which agree well with the
photometric ones. Also, \cite{rod07} compare their spectroscopic
surface gravities with those derived from the stellar positions in
HR diagrams, and reported excellent agreement between the two
surface gravity scales, except for HD 204075, which is the most
massive and brightest star of their sample. We may then accept the
sets of atmospheric parameters of these two works as consistent
and can directly compare the abundances.

The LTE spectrum synthesis calculations were performed by
employing the code by \citet{Spi67} and subsequent improvements in
the past thirty years, which are described in \citet{cay91} and
\citet{barb03}. The adopted model atmospheres (NMARCS) were
computed with a version of the MARCS code, initially developed by
\citet{gben75} and subsequently updated by \citet{plez92}. The
usual notations $\log\epsilon$(A) = $\log(N_A/N_H)$+12 and [A/B] =
$\log(N_A/N_B)_\ast$-$\log(N_A/N_B)_\odot$ were adopted, where
$N_A$ and $N_B$ are the number abundances of species ``A'' and
``B'', respectively. The sources of solar abundances are shown at
the end of Table \ref{medxfe}.

\subsection{Light elements}\label{ablight}

Strong CH and CN bands are characteristics of barium stars, and
these bands are stronger for giants than for dwarfs. Because they
affect the blue region of the spectrum where many lines of heavy
elements appear, it is very important to determine the C and N
abundances before those of heavy elements, especially for giant
stars. The spectrum synthesis of the heavy elements was carried
out only after the C and N abundances were determined. Table
\ref{abun} shows the lines used as reference for each relevant molecule,
C$_2$, CH, and CN, to derive the average of C and N abundances for
the sample stars, which we show in Table \ref{medxfe}. One of the CN
molecule transition suffers from a blend with some Li lines 
around $\lambda$6708, and therefore we had to calculate the Li abundances.
For the stars of the present work, low Li
abundances are expected given that all of them have already
evolved to the red giant branch, where the Li depletion is considerable.
The coolest star HD 220009 is the most Li-poor, with $T_{eff}$ =
4370K and $\log$(Li) = $-$0.94. These results agree 
with Fig. 4 of \citet{di06a}.

Oxygen abundances were derived to be used below in a comparison
with the abundances of elements produced by the $r$-process, since
they are thought to share the same nucleosynthetic origin, massive
stars. The forbidden lines at $\lambda$6300 and $\lambda$6364 were
used, and the results are shown in Tables \ref{abun} and
\ref{medxfe}.

%______________________________________________________________

\subsection{Iron-peak elements}\label{abiron}

For lines of \ion{Mn}{i}, the hyperfine structure ($hfs$) was
taken into account by employing a code made available by Andrew
McWilliam following the calculations described by \citet{proc00}.
The $hfs$ constants for \ion{Mn}{i} were taken from \citet{brod87}
and \citet{walther62}. The nuclear spin (I=2.5) of the only
nuclide that contributes to the manganese abundance ($^{55}$Mn)
was found in \citet{wood57}. Table \ref{hfsMnCu} shows the $hfs$
components, excitation potential and adopted total $\log gf$
values with their references. Abundances derived from the five
available lines of Mn usually agree well. All values of
[Mn/Fe] for the stars of the present work are below solar, as
shown in Fig. \ref{12356fg1}. These low values for [Mn/Fe] for
[Fe/H] $<$ 0.0 agree with the results of
\citet{feltzing07} for disk stars.

For three lines of \ion{Cu}{i}, the $hfs$ from \citet{bi76} was taken
into account, considering the isotopic fractions of 0.69 for
$^{63}$Cu and 0.31 for $^{65}$Cu. In this case, small corrections
were applied in a way that the total $\log gf$ values equal
those of \citet{Bi75} or the National Institute of
Standards \& Technology \citep[NIST,][]{mart02}, which we adopt.
The $hfs$ components for \ion{Cu}{i} lines are listed in Table
\ref{hfsMnCu}. The lines for which $hfs$ were used were checked
using the solar \citep{kurucz84} and Arcturus spectra
\citep{hink00}. Similarly to the report of \citet{di06a}, for some
stars the line $\lambda$5218.2 of \ion{Cu}{i} results in higher
abundances than $\lambda$5105.5 and $\lambda$5782.1.

Four lines of \ion{Zn}{i} were used to compute the Zn
abundances, and usually good agreement was secured for the
abundances resulting from these lines. A difference ranging from
0.25 to 0.6 dex, usually between the $\lambda$4680.1 and
$\lambda$6362.3 lines was observed for two stars of the
present sample, and we note that a similar effect was also
observed in six stars of the \citealt{di06a} sample.

As seen in Fig. \ref{12356fg2}, the abundances for Mn, Cu, and Zn of
\citet{rod07} are larger than in our work, the total $\log
gf$ adopted by the former being the larger. We share four \ion{Mn}{i}
and two \ion{Cu}{i} lines with \citet{rod07}. As they used
larger $\log gf$, lower values for abundances would be expected,
but the opposite occurred. This offset could be caused by the
partially different line set, where ours is the more
extensive. For Zn, \citet{rod07} used only one line, whose 
$\log gf$ is 0.263 lower than ours. The differences
between their Zn abundances and ours are around 0.3
dex, except for two stars, for which an agreement
commensurate with the difference between the $\log gf$'s
is found.

%______________________________________________________________

\subsection{Neutron capture elements}\label{abheavy}

In the present work we only discuss the correlations involving the
abundances of Mn, Cu, and Zn with those of Sr, Y, Ba, Nd 
as representatives of the $s$-process, and Eu, Gd and Dy 
as representatives of the $r$-process, because the nucleosynthetic
production of Ba, Y, and Nd is highly dominated by the
$s$-process by $\sim$81\%, $\sim$92\% and $\sim$65\% respectively,
whereas the $r$-process contributes with $\sim$94\% to the Eu and
$\sim$85\% to the Gd and Dy production, according to
\citet{arlandini99}.
A full analysis for Zr, Mo, Ru, La, Ce, Pr, Sm,
Hf, Pb as well is postponed to a forthcoming paper.
The abundance results for these elements confirm the preliminary
abundance derivations from equivalent widths performed by
\citet{rod07}, where the stars HD 13611, HD 20894, and HD 220009
were found to be normal instead of mild barium stars, given that
the abundance of elements produced by $s$-process is much lower
for these stars than for the other eight that could be called bona
fide barium stars (see Table \ref{medxfe}). The average
abundances of  Sr and Zr shown in Table \ref{medxfe} were computed
only from the lines of \ion{Sr}{ii} and \ion{Zr}{ii}. 
The same $hfs$ employed in \citet{di06a} was used
for all lines of Ba, La, and Eu.

%
%______________________________________________________________

\subsection{Uncertainties}

Uncertainties on abundances were calculated by using an equation
similar to Eq. 14 of \citet{di06a} or Eq. 1 of \citet{digu07}, but
adding a term that takes care of the signal-to-noise (S/N) ratio. New
uncertainties for the abundances found in \citeauthor{di06a} and
\citeauthor{digu07} were derived with Eq. \ref{erapinst} and are given 
in Table \ref{medxfe} along with abundances and uncertainties for
the elements studied here. In these derivations, we used the typical 
uncertainties on the atmospheric parameters determined by \citeauthor{di06a},
$\sigma_{Teff}$ = 100K, $\sigma_{[Fe/H]}$ = 0.04 dex (if $\log g \geq$ 3.3) 
or $\sigma_{[Fe/H]}$ = 0.18 dex (if $\log g <$ 3.3), $\sigma_{\log g}$ = 0.1 dex,
and $\sigma_\xi$ = 0.1 km/s (see their Sect. 3.5), and those  
by \citet{rod07}, $\sigma_{Teff}$ = 50 K, $\sigma_{[Fe/H]}$ = 0.1 dex 
$\sigma_{\log g}$ = 0.35 dex, and $\sigma_\xi$ = 0.06 km/s 
(see Table \ref{patm}).

Recalling that ``$Ap$'' is  the output of the synthesis
program, we have

{\scriptsize
\begin{equation}
\label{erapinst} \sigma_{Ap}=\sqrt{(\Delta A_T)^2+(\Delta
A_{mt})^2+(\Delta A_l)^2+(\Delta A_\xi)^2+(\Delta A_{sn})^2},
\end{equation}}

\noindent where $\Delta A_T$, $\Delta A_{mt}$, $\Delta A_l$,
$\Delta A_\xi$ are the differences on $A_p$ for each
element because of variations of 1$\sigma$ in the temperature,
metallicity, $\log g$, and microturbulence velocity,
respectively. The last term, $\Delta A_{sn}$, is the difference on
$A_p$ after adding (S/N)$^{\rm -1}$ to the spectrum of the
reference stars defined in \citeauthor{di06a} and
\citeauthor{digu07}. The S/N ratios shown in Table \ref{signoise}
were measured in a window around $\lambda$6000.

The variations on the abundances owing to the atmospheric
parameters were computed for only one line, and we extended this result
to all other lines of a given element. Indeed, the
ideal procedure would be to compute all variations for all lines,
since each line reacts differently to these variations, according
to their atomic constants. This process entails
synthesizing each line more than five times. This
very time consuming approach does not compare favorably with a
simpler procedure, when one considers the resulting very
small improvement in accuracy. We were nevertheless very
careful in the choice of the line employed for the uncertainty
evaluation, to minimize the loss in accuracy by adopting
the simpler procedure. The profile of the adopted
line must be as well defined as possible, neither too
strong nor too weak, and also affected by noise in an
average way, to make the line useful.

By adopting this procedure, the larger the number of lines
employed to derive the abundance of a given element, the
lower the expected loss of accuracy. Thus, $n$ in
Eq. 15 of \citeauthor{di06a} is the number of lines used to
compute the average of the abundance for each element for each
star, shown in their Table 15.

%------------Table signoise-------------------------------------------------------------------------------------
\begin{table}
\scriptsize{ \caption{Signal to noise measured around $\lambda$6000 
for stars analyzed here and in \citet{di06a}.} 
\label{signoise} 
\centering
\setlength\tabcolsep{2pt}
\begin{tabular}{lclclclc}
\noalign{\smallskip}
\hline \hline
Star & SN & Star & SN & Star & SN & Star & SN \\
\hline
\noalign{\smallskip}

HD 749    & 220 & HD 26967* & 180 & HD 107574  & 230 & HD 188985  & 150 \\
HD 2454   & 210 & HD 27271  & 220 & HD 113226* & 330 & HD 202109  & 190 \\
HD 5424   & 150 & HD 46407  & 160 & HD 116713  & 110 & HD 204075  & 320 \\
HD 8270   & 230 & HD 48565  & 220 & HD 116869  & 190 & HD 205011  & 210 \\
HD 9362*  & 310 & HD 76225  & 140 & HD 123396  & 200 & HD 210709  & 170 \\
HD 12392  & 180 & HD 87080  & 200 & HD 123585  & 100 & HD 210910  & 260 \\
HD 13551  & 190 & HD 89948  & 370 & HD 139195  & 420 & HD 220009* & 390 \\
HD 13611* & 370 & HD 92545  & 140 & HD 147609  & 150 & HD 222349  & 130 \\
HD 20894* & 440 & HD 104979 & 140 & HD 150862  & 180 & BD+18 5215 & 100 \\
HD 22589  & 190 & HD 106191 & 110 & HD 181053  & 240 & HD 223938  & 360 \\
\noalign{\smallskip}
\hline
\end{tabular}
}
\end{table}
%----------------------------------------------------------------------------------------------

\subsection{Abundance scatter}

Abundance ratios from our sample and stars from the
literature are shown in Figs. \ref{12356fg1} and 
\ref{12356fg3} - \ref{12356fg8}. In general, the [Mn/Fe] 
derived by us follows the
trend found in the literature, except for the high values of some
stars from \citet{lb91} and \citet{scl93}. The stars
with the lowest values for [Cu/Fe] in Fig. \ref{12356fg1} are those
with metallicities and kinematics characteristic of the halo, with
[Fe/H] $< -$1. Some of these stars, indicated with a big open
square in Figs. \ref{12356fg1}, \ref{12356fg3}, \ref{12356fg4},
\ref{12356fg5}, and \ref{12356fg8}, figure in \citet{menn97}'s work,
and were classified by them as belonging to the halo population
through kinematical criteria. It is worthwhile to note, however,
that the abundance patterns of some of these stars do not match 
\citeauthor{menn97}'s kinematical criteria. 
Furthermore, the [Cu/Fe] value
of He2-467 seems too low to be explained away by membership in the
halo population of the Galaxy, whereas some stars from
\citet{scl93} have [Cu/Fe] too high for disk stars. Yet [Zn/Fe]
from all works in Fig. \ref{12356fg1} follows the trend found in
the literature in this range of metallicities.

High scatter may prevent accurate conclusions about abundance
trends. Some degree of scatter
is always expected, but the origin of such a high degree of
scatter as seen in Fig. \ref{12356fg1} for Mn abundances is
elusive. Intrinsic cosmic scatter can be the result
of the heterogenous nature of the original interstellar medium
from which each star was formed. However, all too often
scatter arises from differences in the procedure used by different
authors, which could involve parameters as divers as different 
line sets, atomic
constants, $hfs$ employed \citep{delpelosoetal2005}, 
solar abundances as reference, atmosphere models, and
spectrum synthesis codes.

Table \ref{lit} shows that the difference between results
obtained for Mn abundances by different authors for some stars can
reach 0.50 dex, whereas for Cu and Zn this difference is smaller.
For these eight stars, only the abundances from this work and 
\citet{di06a} are shown in figures. The different values for $\log gf$
adopted by different authors might be one of the main sources of
the scatter in the abundance ratios, since an increase 
of a certain amount in the $\log gf$
results in a reduction of approximately the same amount in the
abundance \citep{gray92}. We thus preferred to use in the fits
only the abundances from the literature derived from the laboratory 
$\log gf$s. Yet, this cannot allow a conclusion as to why the
anticorrelation suggested by \citet{cps99} between [Cu/Fe] and
[Ba/Fe] is not verified in our analysis (Fig. \ref{12356fg3}), 
but we note that the abundance interval considered here 
is larger than in \citet{cps99}, and our
conclusions consequently more robust.

Considering the binary scenario for the origin of stars that show
enhanced $s$-process elements, and their appearance in any
galactic population, one might suggest that the origin of at least
part of the scatter in the case of $s$-process enriched stars
could reside in the different composition of the material
dredged up to the surface of the former primary component and
transferred onto the secondary component. This is supported by
Figs. \ref{12356fg3} - \ref{12356fg5}, where the
scatter seems to be larger when the Mn, Cu, and Zn abundances are
related to those of an element with a larger contribution of the
$s$-process for its formation, namely Ba, Y, and Nd, while an
apparent low scatter is found when the abundances are related to
those of the $r$-process elements, Eu, Gd, and Dy. We note,
however, that data for Eu, Gd, and Dy are much scarcer in the
literature compared to Ba, Y, and Nd, which are more widely studied
and are therefore more likely to involve differences in method.
Indeed, we employed for Dy only data from \citet{di06a} and our work, 
which both employ the same atomic data and spectral synthesis routine.

The main $s$-processing strongly depends on the initial mass 
of the AGB star \citep[see e.g.][]{bisterzo10}. A TP-AGB with low initial 
mass suffers less third dredge-up episodes than stars with 
higher masses \citep[e.g.][]{lugaro03,bisterzo10}. 
Therefore, the degree of enrichment in carbon and heavy elements of the 
barium star depends on the initial mass of the companion white dwarf. 
\citet[][and references therein]{chen03} studied how 
to recover the initial mass of the white dwarf in binary systems.

Furthermore, the transfer of enriched gas from the AGB star to the 
pre-barium star depends on their masses and the binary separation.
Because the AGB star wind is generally much stronger than that of the 
pre-barium star, the wind of the latter is usually neglected 
in the calculations of mass-loss/accretion rate through winds in binary systems 
\citep{han95}. The higher the AGB mass, the larger its mass-loss rate 
and the higher its wind velocity. A dependence of the wind velocity of the 
primary star on its radius and the binary separation is given by \citet{kool95}. 
High AGB wind velocities result in lower mass accretion 
rate by the pre-barium star. On the other hand, the mass accretion rate undergone by the 
pre-barium star directly depends on its mass. A large separation of 
the two components of the binary system results in a higher wind 
velocity that reductes the mass accretion rate of the pre-barium star.

%
%______________________________________________________________

\section{Discussion}\label{disc}

\citet{cps99} have suggested an anticorrelation between [Cu/Fe]
and [Ba/Fe] for a sample of solar-type disk stars, including stars
from the young, solar-metallicity Ursa Major kinematical group:
this anticorrelation was proposed before by \cite{pp97} and
\cite{psc98} for two Ba-enriched symbiotic stars. The [Cu/Fe] vs.
[Ba/Fe] run including stellar abundances of the present work and
others selected from the literature shows no evidence of such an
anticorrelation, as can be clearly seen in Fig. \ref{12356fg3}b. The
pattern seen in Fig. 6 of \citeauthor{cps99} can indeed be seen in
our Fig. \ref{12356fg3}b, but only by isolating the data used in their
plot. Indeed, their results along with ours and those of
\citet{di06a} essentially follow a flat trend. 
With the aim to more extensively investigate this suspected
anticorrelation, we included other iron peak elements (Mn and Zn)
and verified their behavior as it relates to Ba and other 
$s$- and $r$-elements.

%---------Table lit------------------------------------------------------------------------------------
\onltab{6}{
\begin{table*}
\caption{Data from the literature for stars that were analyzed
in more than one work.} 
\label{lit} 
\centering
\begin{tabular}{llrrrrrrrrrr}
\noalign{\smallskip}
\hline \hline
Star & Ref & [Fe/H] & [Mn/Fe] & [Cu/Fe] & [Zn/Fe] & [Y/Fe] & [Ba/Fe] & [Nd/Fe] & [Eu/Fe] & [Gd/Fe] & [Dy/Fe] \\
\hline
\noalign{\smallskip}
HD 8270   & 1 & -0.42 &  -0.29 & -0.21 & -0.04 &  0.95 &  1.11 &  0.73 &  0.32 &  0.25 &  0.04 \\
HD 8270   & 2   & -0.53 &  -0.07 & -0.07 &  0.04 &  0.75 &  1.17 &  0.80 &  0.19 &  ...  &  ...  \\
\hline
\noalign{\smallskip}
HD 13551  & 1 & -0.44 &  -0.30 & -0.17 &  0.10 &  1.08 &  1.16 &  0.73 &  0.21 &  0.55 &  0.09 \\
HD 13551  & 2   & -0.28 &  -0.22 &  0.03 &  0.04 &  0.90 &  1.38 &  0.53 &  ...  &  ...  &  ...  \\
\hline
\noalign{\smallskip}
HD 22589  & 1 & -0.27 &  -0.10 &  0.00 &  0.11 &  0.83 &  0.88 &  0.32 &  0.21 &  0.03 &  0.04 \\
HD 22589  & 2   & -0.16 &   0.16 &  0.06 &  0.06 &  0.72 &  0.75 &  0.07 &  0.26 &  ...  &  ...  \\
\hline
\noalign{\smallskip}
HD 87080  & 1 & -0.44 &  -0.31 & -0.25 &  0.10 &  1.11 &  1.48 &  1.56 &  0.66 &  0.90 &  1.14 \\
HD 87080  & 3  & -0.51 &  -0.09 &  0.01 &  0.26 &  1.01 &  1.51 &  0.97 &  0.61 &  ...  &  ...  \\
\hline
\noalign{\smallskip}
HD 89948  & 1 & -0.30 &  -0.17 & -0.17 &  0.02 &  1.02 &  0.99 &  0.65 &  0.16 &  0.25 & -0.31 \\
HD 89948  & 4 & -0.27 &   0.18 &  0.08 &  ...  &  1.11 &  0.83 &  0.60 &  ...  &  ...  &  ...  \\
HD 89948  & 5  & -0.12 &  -0.32 &  ...  & -0.24 &  0.85 &  0.86 &  0.59 &  ...  &  ...  &  ...  \\
\hline
\noalign{\smallskip}
HD 123585 & 1 & -0.48 &  -0.30 &  0.04 &  0.10 &  1.34 &  1.79 &  1.41 &  0.83 &  0.55 &  0.64 \\
HD 123585 & 5  & -0.50 &   ...  &  ...  &  ...  &  1.14 &  1.32 &  0.98 &  ...  &  ...  &  ...  \\
\hline
\noalign{\smallskip}
HD 150862 & 1 & -0.10 &  -0.14 & -0.12 &  0.00 &  1.08 &  1.03 &  0.34 &  0.20 &  0.23 &  0.14 \\
HD 150862 & 5  & -0.22 &   0.22 &  ...  & -0.25 &  0.65 &  ...  &  0.32 &  ...  &  ...  &  ...  \\
\hline
\noalign{\smallskip}
HD 202109 & 6  & -0.04 &  -0.35 & -0.17 & -0.22 &  0.44 &  0.57 &  0.18 &  0.17 &  0.06 & -0.05 \\
HD 202109 & 7   &  0.01 &  -0.25 & -0.01 & -0.08 & 0.48  & ...   & 0.42  & 0.32  & ...   &  0.33 \\
\noalign{\smallskip}
\hline
\end{tabular}

\tablebib{
(1) \citet{di06a}; (2) \citet{p05};
(3) \citet{pj03}; (4) \citet{scl93}; (5) \citet{lb91};
(6) This; (7) \citet{yush04}.
}
\end{table*}
}
%---------------------------------------------------------------------------------------------

%------------Table fits----------------------------------------------------------------------------------------
\onltab{7}{
\begin{table*}
\caption{Results of the least-squares fits for Figs. \ref{12356fg3}, \ref{12356fg4},
\ref{12356fg5}, and \ref{12356fg8}.
``Cov'' is the covariance between ``A'' and ``B'' for the 
linear fit Y(X) = AX + B (first column), 
and ``D.O.F.'' is the number of degrees of freedom.} 
\label{fits} 
\centering
\setlength\tabcolsep{2pt}
\begin{tabular}{lllr}
\noalign{\smallskip}
\hline \hline
\noalign{\smallskip}
\multicolumn{4}{c}{Least-squares fits for Fig. \ref{12356fg3}} \\
Y(X) = AX + B & $\chi^2_{red}$ & Cov & D.O.F \\
\noalign{\smallskip}
\hline
\noalign{\smallskip}
\hbox{[Mn/Fe]} = -0.030(40)[Ba/Fe] - 0.220(50) & 2.616 & -0.001820 & 33 \\
\hbox{[Cu/Fe]} = 0.045(50)[Ba/Fe] - 0.200(60)  & 1.072 & -0.003206 & 33 \\
\hbox{[Zn/Fe]} = 0.160(60)[Ba/Fe] - 0.195(70)  & 1.748 & -0.004097 & 33 \\
\hbox{[Mn/Fe]} = 0.060(40)[Y/Fe] - 0.310(40)   & 2.564 & -0.001784 & 33 \\
\hbox{[Cu/Fe]} = 0.120(60)[Y/Fe] - 0.260(60)   & 0.955 & -0.003019 & 33 \\
\hbox{[Zn/Fe]} = 0.250(70)[Y/Fe] - 0.240(60)   & 1.513 & -0.004019 & 33 \\
\noalign{\smallskip}
\hline \hline
\noalign{\smallskip}
\multicolumn{4}{c}{Least-squares fits for Fig. \ref{12356fg4}} \\
& $\chi^2_{red}$ & Cov & D.O.F \\
\noalign{\smallskip}
\hline
\noalign{\smallskip}
\hbox{[Mn/Fe]} = -0.052(30)[Nd/Fe] - 0.210(20) & 2.512 & -0.000589 & 33 \\
\hbox{[Cu/Fe]} = 0.010(40)[Nd/Fe] - 0.150(30)  & 1.095 & -0.001122 & 33 \\
\hbox{[Zn/Fe]} = 0.185(40)[Nd/Fe] - 0.160(40)  & 1.371 & -0.001469 & 33 \\
\hbox{[Mn/Fe]} = -0.090(60)[Eu/Fe] - 0.220(20) & 2.185 & -0.001024 & 38 \\
\hbox{[Cu/Fe]} = 0.040(60)[Eu/Fe] - 0.170(20)  & 1.893 & -0.001006 & 38 \\
\hbox{[Zn/Fe]} = 0.370(90)[Eu/Fe] - 0.140(30)  & 1.605 & -0.002642 & 38 \\
\noalign{\smallskip}
\hline \hline
\noalign{\smallskip}
\multicolumn{4}{c}{Least-squares fits for Fig. \ref{12356fg5}} \\
& $\chi^2_{red}$ & Cov & D.O.F \\
\noalign{\smallskip}
\hline
\noalign{\smallskip}
\hbox{[Mn/Fe]} = -0.000(40)[Gd/Fe] - 0.250(20) & 2.286 & -0.000596 & 37 \\
\hbox{[Cu/Fe]} = 0.055(50)[Gd/Fe] - 0.180(30)  & 1.008 & -0.001135 & 37 \\
\hbox{[Zn/Fe]} = 0.290(70)[Gd/Fe] - 0.140(40)  & 1.047 & -0.001965 & 37 \\
\hbox{[Mn/Fe]} = -0.050(30)[Dy/Fe] - 0.230(20) & 2.230 & -0.000323 & 36 \\
\hbox{[Cu/Fe]} = 0.025(40)[Dy/Fe] - 0.160(20)  & 1.070 & -0.000630 & 36 \\
\hbox{[Zn/Fe]} = 0.180(50)[Dy/Fe] - 0.090(25)  & 1.721 & -0.000877 & 36 \\
\noalign{\smallskip}
\hline \hline
\noalign{\smallskip}
\multicolumn{4}{c}{Least-squares fits for Fig. \ref{12356fg8}} \\
& $\chi^2_{red}$ & Cov & D.O.F \\
\noalign{\smallskip}
\hline
\noalign{\smallskip}
\hbox{[Mn/Fe]} = 0.050(40)[Sr/Fe] - 0.300(40) & 2.548 & -0.001253 & 32 \\
\hbox{[Cu/Fe]} = 0.050(50)[Sr/Fe] - 0.190(40) & 1.017 & -0.002009 & 32 \\
\hbox{[Zn/Fe]} = 140(60)[Sr/Fe] - 0.130(50)   & 1.728 & -0.002768 & 32 \\
$\log\epsilon$(Mn) = 1.030(80)$\log\epsilon_w$(Sr) + 3.020(140)  & 0.952 & -0.010765 & 37 \\
$\log\epsilon$(Cu) = 0.990(90)$\log\epsilon_w$(Sr) + 1.990(160)  & 0.690 & -0.014851 & 37 \\
$\log\epsilon$(Zn) = 0.620(70)$\log\epsilon_w$(Sr) + 3.180(10)   & 1.867 & -0.008325 & 37 \\
$\log\epsilon$(Mn) = -0.020(60)$\log\epsilon_m$(Ba) + 4.840(170) & 6.992 & -0.010044 & 32 \\
$\log\epsilon$(Cu) = 0.100(70)$\log\epsilon_m$(Ba) + 3.410(210)  & 4.615 & -0.014135 & 32 \\
$\log\epsilon$(Zn) = 0.010(50)$\log\epsilon_m$(Ba) + 4.200(150)  & 4.908 & -0.007749 & 32 \\

\noalign{\smallskip}
\hline
\end{tabular}
\end{table*}
}
%----------------------------------------------------------------------------------------------

\subsection{Relations with heavy elements}

In order to evaluate the correlations between iron peak and
neutron capture elements, least-square fits were performed
considering the uncertainties provided by the authors of the
corresponding abundance analysis, the results of which we show in
Table \ref{fits}. For all fits the covariance is around zero,
indicating that the two variables involved may be independent. If
both ratios are below or above the expected value, the covariance
is positive. When one ratio is above and the other is below the
expected value, the covariance is negative. 

For the fits we
considered only the barium stars analyzed here, 
\citet{di06a} and one star of \citet{pj03}. The same stars were used in the
fits involving Eu in Fig. \ref{12356fg4}.
This is indicated in the labels of Figs. \ref{12356fg3} and
\ref{12356fg4}.

The correlation with Dy is expectedly poor owing to the
low number of available lines of this element. For the star HD
5424 the Dy abundance is unusually high, (Dy/Fe] = 1.65,
\citep[see][]{di06a}. A higher Dy abundance for barium stars is
expected because this element is in the $s$-process path, but an 
abundance as high as this is probably suspect. However, the $\chi^2_{red}$
for [Cu/Fe] vs [Dy/Fe] in Fig. \ref{12356fg5} is very close to 1, indicating 
the good quality of the fit, even if the HD 5424 data point is kept.

The runs involving [Mn/Fe] show very weak correlations and
low quality in the fits, with $\chi^2_{red} \sim$ 2.4. The large
scatter clearly prevents better fits as well as a more definitive
conclusion. The
absent correlation of [Mn/Fe] with the $r$-element abundances is
not compatible with a significant production of Mn in SN II.
Similarly, the lack of correlation with the $s$-element abundances
indicates that the bulk of the Mn production is very probably not
taking place in AGB stars. This indicates that the production of
Mn is a mix of several processes.

Indeed, several works have proposed quite diverse sites for
the nucleosynthesis of Mn. \citet{gratton89} attributed the
significant increase of [Mn/Fe] for [Fe/H] $> -$1 to an
overproduction of Mn by SN Ia, whereas \citet{mcwilliam03}
suggested that this effect could be a result of 
metallicity-dependent yields of Mn in both SN Ia and SN II
\citep[e.g.][]{arnett71,ww95,chieffi04,limongi05}. The Mn yields
of SN II would increase as more metal-rich progenitors produce
higher amounts of Mn, producing the increasing [Mn/Fe] trend
already seen by \citet{gratton89}. Regarding the $s$-process, the
weak component could eventually contribute to the production of
Mn, because this component acts for nuclei with A $<$ 56, whereas no
significant contribution is expected from the main component for
this region of atomic mass. The behavior of Mn with these
components is discussed in Sect. \ref{weakmain}. Our results 
therefore support an interpretation in which the bulk of the 
Mn production originates in SN Ia.

The runs of [Cu/Fe] vs. [Ba,Y,Nd/Fe] show a slightly increasing
trend, even though the slope of the relation with Ba and Y is
clearly of little significance. We found no trend for Nd. The 
runs of [Cu/Fe] vs. [Eu/Fe] also show a
slightly increasing trend with much lower scatter compared to
the $s$-process elements. Particularly, the runs of [Cu/Fe] with
[Gd,Dy/Fe] are fairly tight and essentially flat. The weak
correlation between Cu and the $r$-elements indicates that the
bulk of the Cu production does not come from SN II. According to
the \citet{mish02} estimate, the bulk of the Cu abundance ($\sim$62-65\%)
should be owed to SN Ia, in agreement with \citet{matte93}, 25\%
to a secondary process in massive stars and only a small fraction
(7-8\%) to a primary process. A similar result was obtained by
\citet{cunhaetal2002} for the $\omega$ Cen globular cluster, where
these authors conclude that the $r$-process contribution to the
synthesis of Cu must be low. Our result is thus in line with
previous literature results, which indicate a low contribution of 
SN II to the synthesis of Cu.

The [Zn/Fe] vs. [Ba,Y,Nd/Fe] runs show a clearly increasing trend
with statistically significant correlations. The runs of [Zn/Fe]
vs. [Eu/Fe] show an increasing trend with much lower scatter when
compared to the [Zn/Fe] vs. [Ba,Nd,Y/Fe] plots, and the
correlation due to Zn is stronger than in the case of Cu. [Zn/Fe]
increases much faster than [Cu/Fe] with [Eu,Gd,Dy/Fe]. This result
suggests that for Zn a higher abundance fraction is owed to
$r$-process than for Cu. \citet{mish02} concluded
that $\sim$30\% of the synthesis of Zn would be owing to a primary
source acting in massive stars, similarly to Fe, $\sim$67\% would
originate in SN Ia and only $\sim$3\% of the abundance of Zn could
be owing to AGB stars. The stronger correlation between Zn and
$r$-elements agrees with the conclusion that the
contribution of the $r$-process is more significant for Zn than
for Cu.

%------figure 12356fg1---------------------------------------------------------------------------
\onlfig{1}{
\begin{figure*}
\centering
\includegraphics[width=16cm]{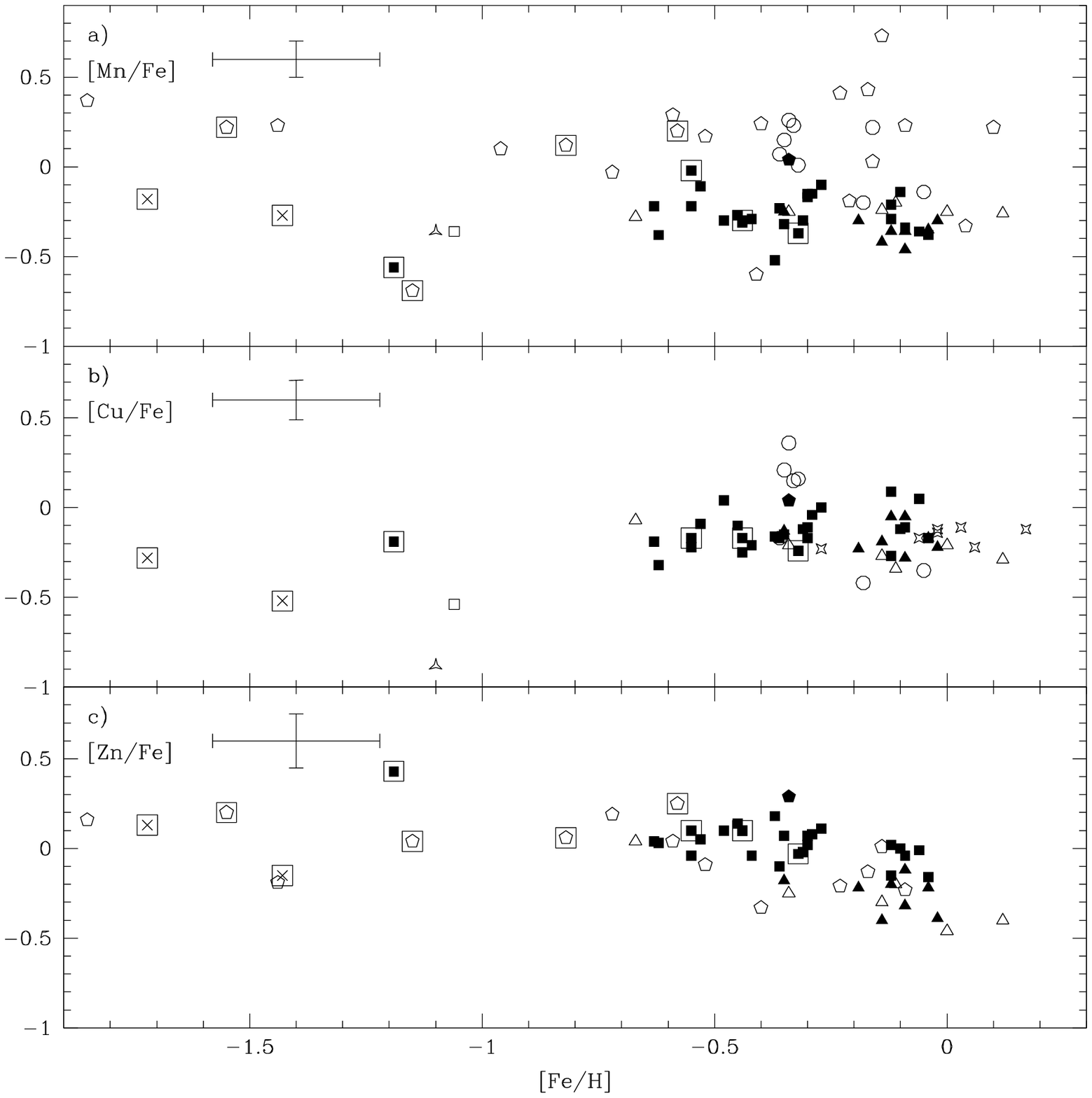}
\caption{[Mn/Fe], [Cu/Fe] and [Zn/Fe] vs. [Fe/H].
Error bars represent the highest values for the uncertainties for each axis.
Symbols:
triangles are results from this work, where filled
triangles are barium stars and open triangles, those
stars considered to be normal by \citet{rod07}; filled squares are data
taken from \citet{di06a};
filled pentagon: \citet{pj03};
crosses: \citet{jp01};
open square: BD-21$^{\circ}$ 3873, a halo symbiotic star of \citet{pp97};
starred triangle: HE2-467, a halo yellow symbiotic star of \citet{psc98};
starred squares: \citet{cps99};
open pentagons: \citet{lb91};
open circles: \citet{scl93}.
Big open squares involving some points indicate the halo barium stars,
according to \citet{menn97}.}
\label{12356fg1}
\end{figure*}
}
%------------------------------------------------------------------------------------------------

%------figure 12356fg2---------------------------------------------------------------------------
\onlfig{2}{
\begin{figure*}
\centering
\includegraphics[width=16cm]{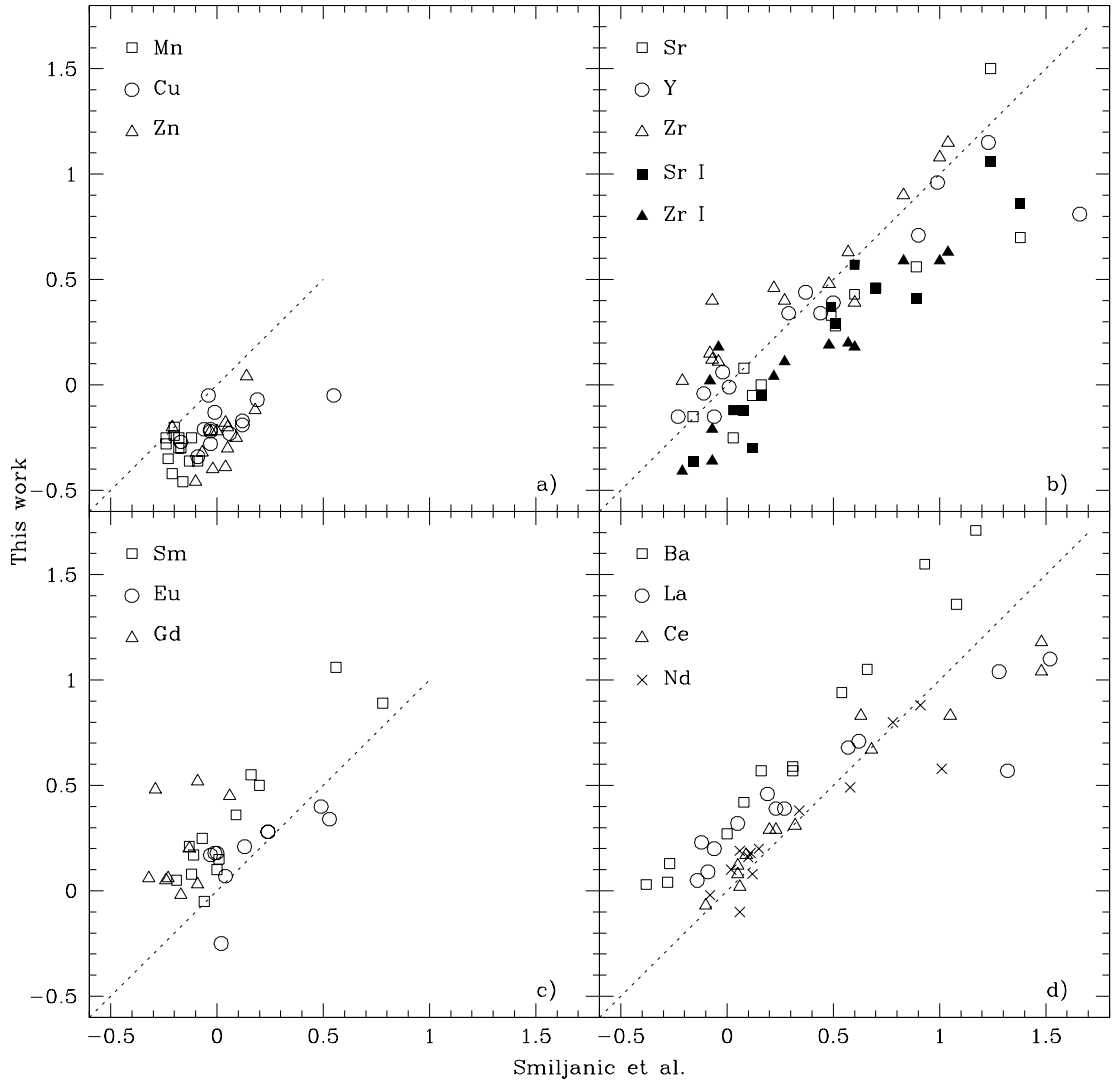}
\caption{Comparison between [X/Fe] of \citet{rod07} and this work. The
dotted lines indicate the same values for both. In panel b) full squares
and the full triangles represent the comparison with the average for lines
of Sr I and Zr I, respectively, of this work, whereas open squares and triangles
represent Sr II and Zr II, respectively.}
\label{12356fg2}
\end{figure*}
}
%---------------------------------------------------------------------------------------------

%------figure 12356fg3---------------------------------------------------------------------------
\onlfig{3}{
\begin{figure*}
\centering
\includegraphics[width=16cm]{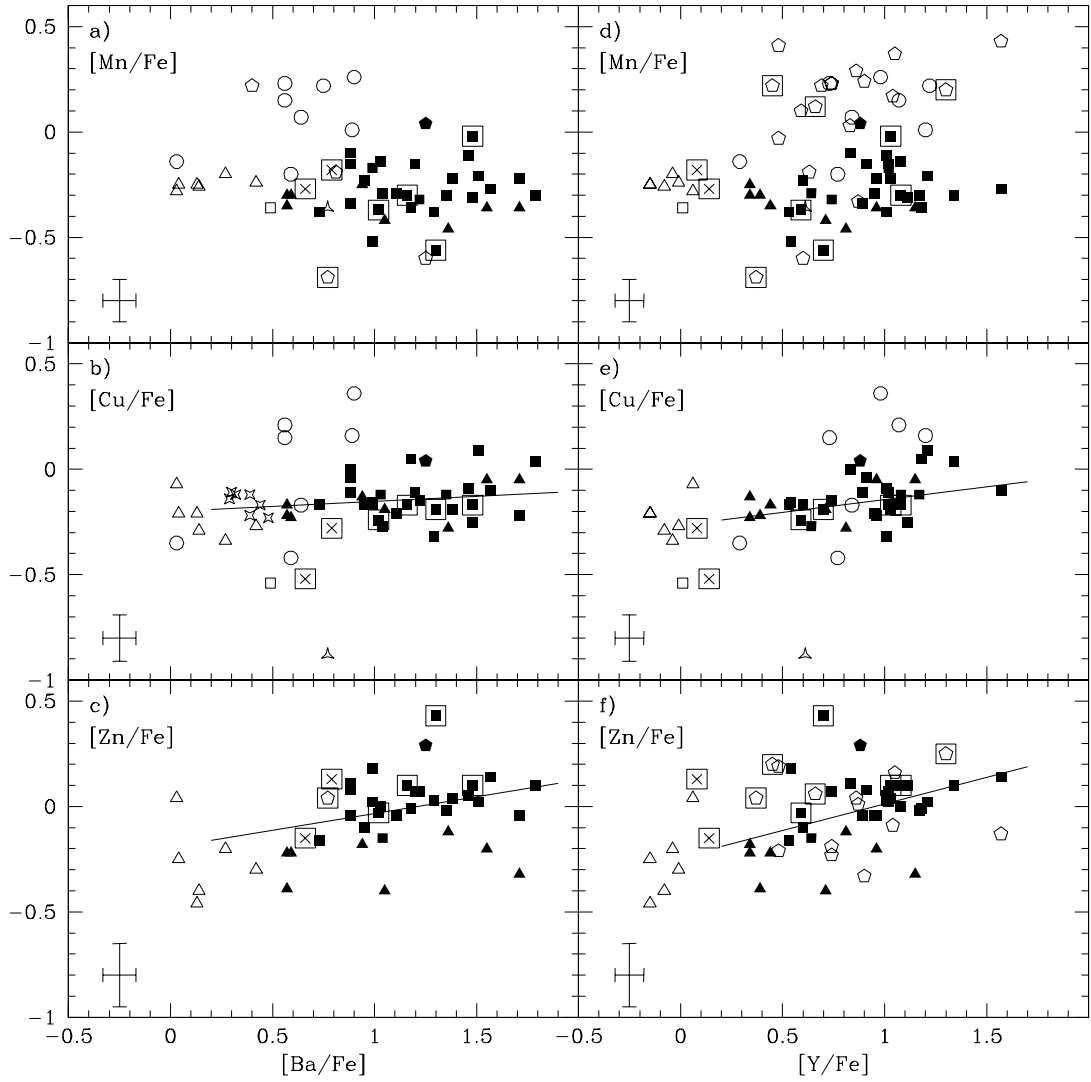}
\caption{Abundance ratios involving Mn, Cu, Zn, Ba, Y, and Fe.
Error bars are defined as in Fig. \ref{12356fg1}. Results of the least-squares fits 
are shown for each panel. Only filled symbols were used in the fits.
The straight line is shown
only for fits with $\chi^2_{red} < $ 2. Symbols are the same as in Fig. \ref{12356fg1}.
The results for the least-squares fits are given in Table \ref{fits}.}
\label{12356fg3}
\end{figure*}
}
%-----------------------------------------------------------------------------------------------

%------figure 12356fg4---------------------------------------------------------------------------
\onlfig{4}{
\begin{figure*}
\centering
\includegraphics[width=16cm]{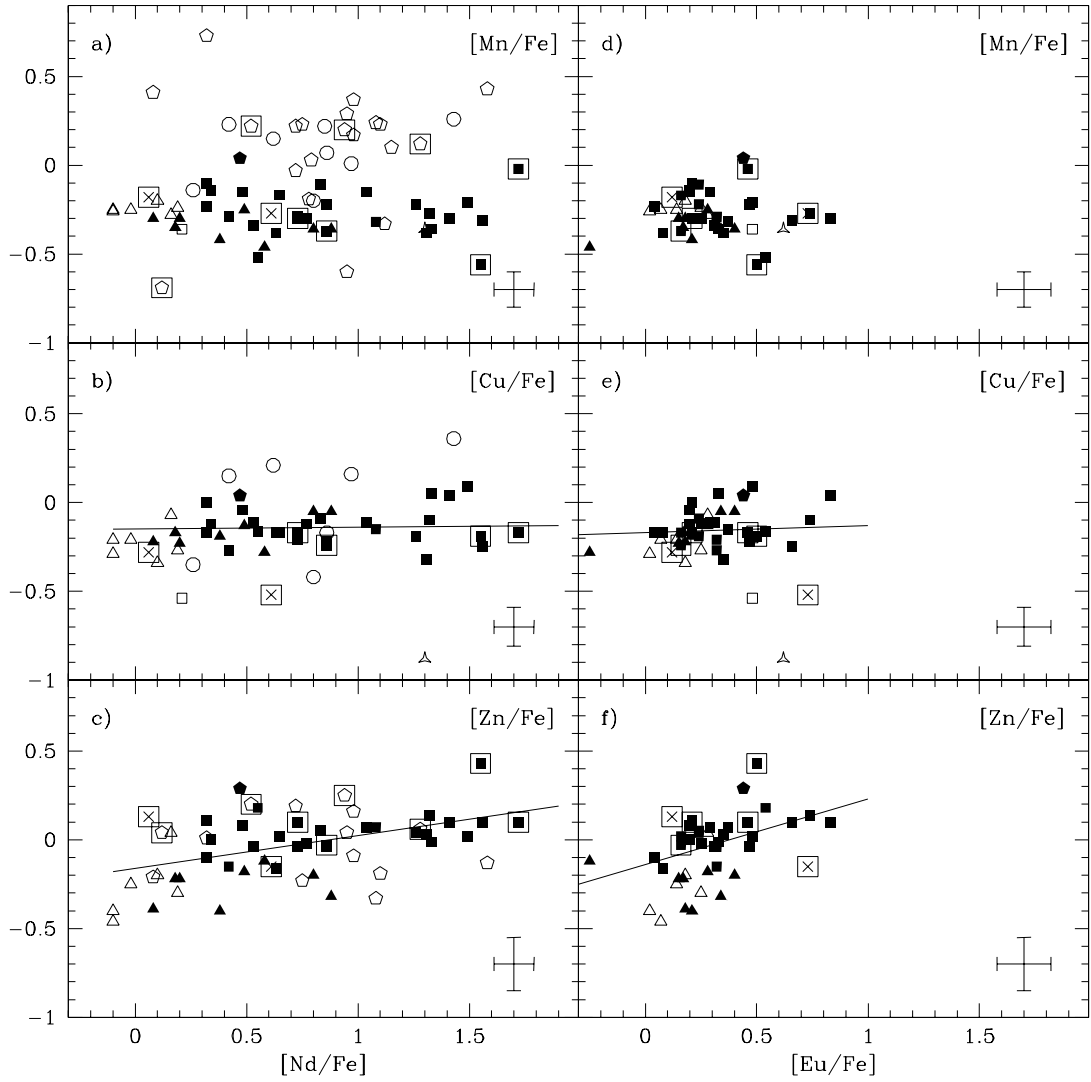}
\caption{Abundance ratios involving Mn, Cu, Zn, Nd, Eu, and Fe.
Error bars are defined as in Fig. \ref{12356fg1}.
Results of the least-squares fits are shown for each panel. For the fits a), b) and c),
only filled symbols were used, and for d), e) and f), normal stars (open triangles) were
included, except HD 113226.
The straight line is shown only for fits with $\chi^2_{red} < $ 2.
Symbols are the same as in Fig. \ref{12356fg1}.
The results for the least-squares fits are given in Table \ref{fits}.}
\label{12356fg4}
\end{figure*}
}
%----------------------------------------------------------------------------------------------

%------figure 12356fg5---------------------------------------------------------------------------
\onlfig{5}{
\begin{figure*}
\centering
\includegraphics[width=16cm]{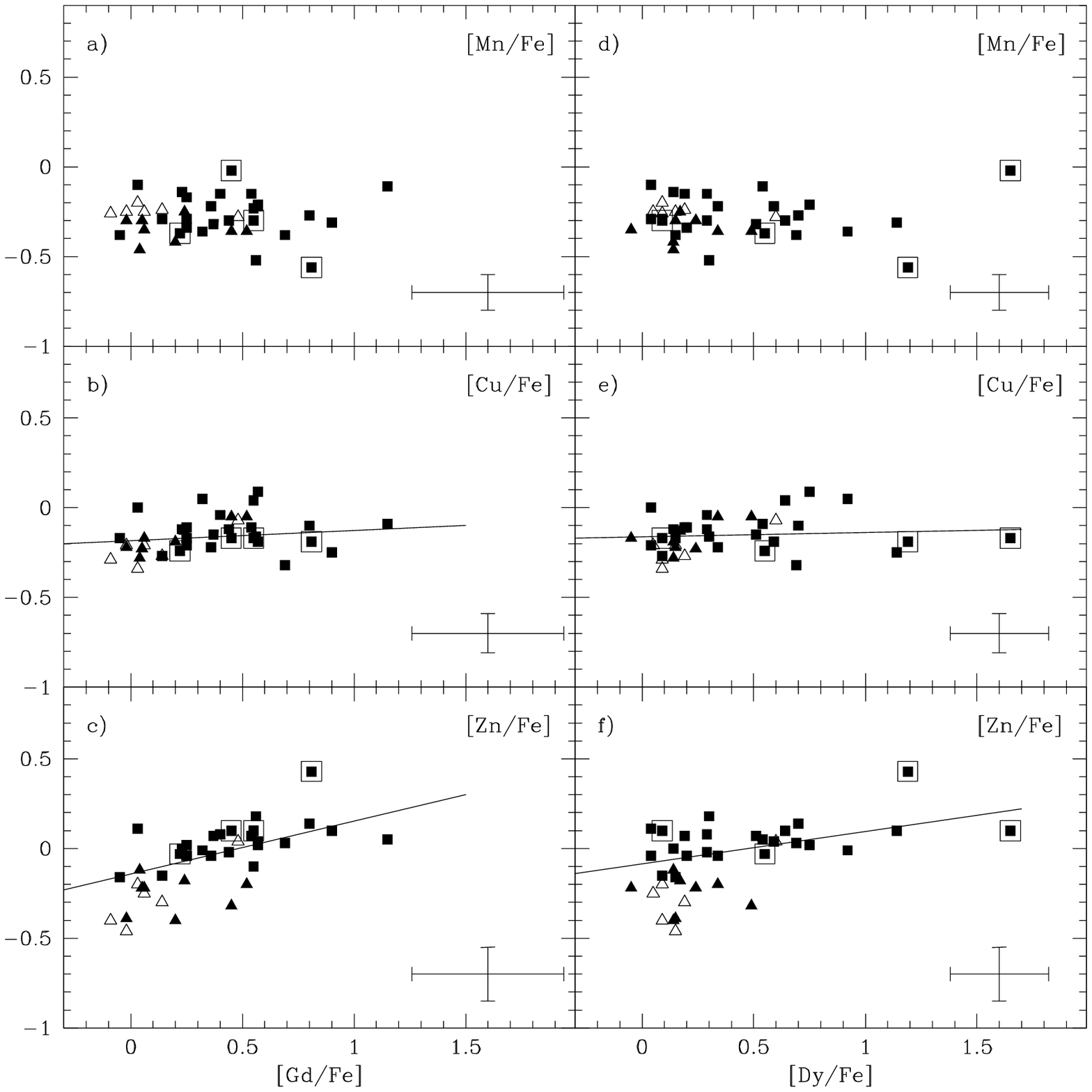}
\caption{Abundance ratios invoving Mn, Cu, Zn, Gd, Dy, and Fe.
Error bars are defined as in Fig. \ref{12356fg1}.
Symbols: triangles are results of this work, where filled are barium stars
and open, those stars considered to be normal instead of barium stars by \citet{rod07};
Filled squares: \citet{di06a}.
Results of the least-squares fits are shown for each panel. Only filled symbols were
used in the fits. The straight line is shown only for fits with $\chi^2_{red} < $ 2.
The results for the least-squares fits are given in Table \ref{fits}.}
\label{12356fg5}
\end{figure*}
}
%----------------------------------------------------------------------------------------------

%------figure 12356fg6---------------------------------------------------------------------------
\begin{figure}
\centering
\includegraphics[width=9cm]{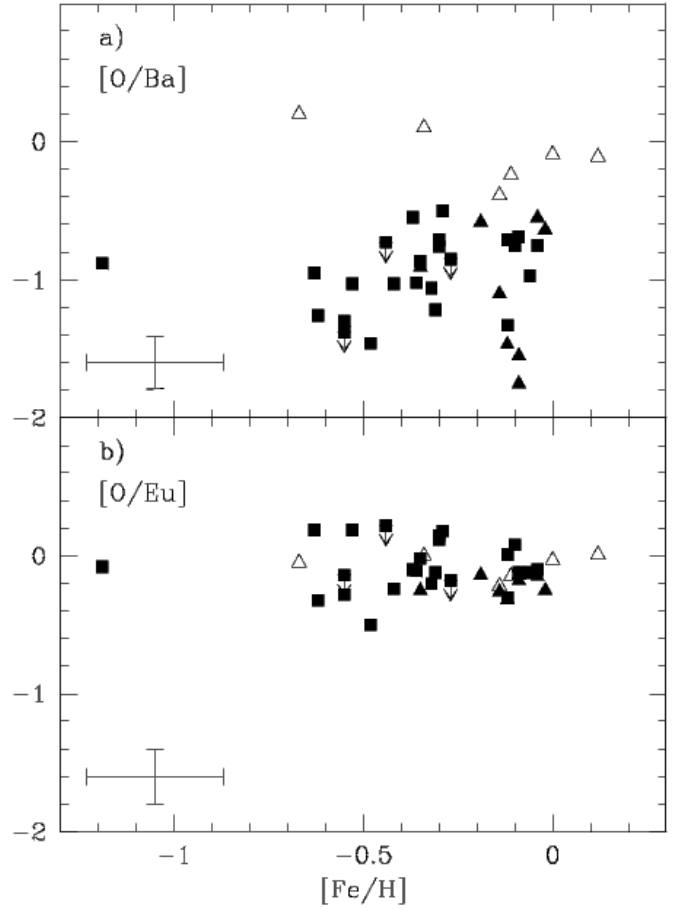}
\caption{[O/Ba] and [O/Eu] vs. [Fe/H]. Symbols are the same as in
Fig. \ref{12356fg7}. Error bars are defined as in Fig. \ref{12356fg1}.}
\label{12356fg6}
\end{figure}
%---------------------------------------------------------------------------------------------

%------figure 12356fg7---------------------------------------------------------------------------
\onlfig{7}{
\begin{figure*}
\centering
\includegraphics[width=16cm]{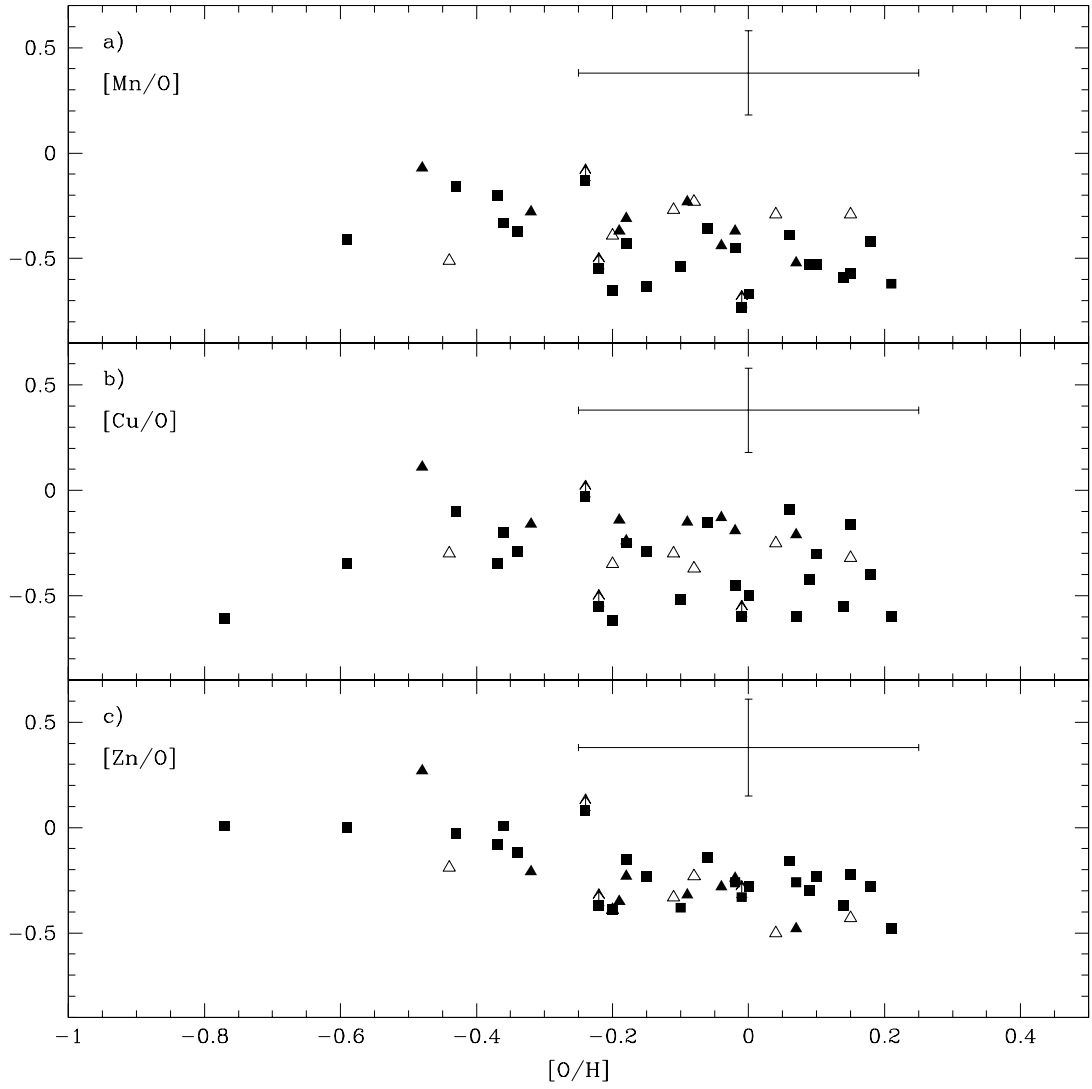}
\caption{[Mn/O], [Cu/O] and [Zn/O] vs. [O/H].
Error bars are defined as in Fig. \ref{12356fg1}.
Arrows up indicate lower limits for the ratios.
Symbols: triangles are results of this work, where filled are barium stars and open,
those stars considered to be normal instead of barium stars by
\citet{rod07}; filled squares are data taken from \citet{di06a}.}
\label{12356fg7}
\end{figure*}
}
%---------------------------------------------------------------------------------------------

%------figure 12356fg8---------------------------------------------------------------------------
\onlfig{8}{
\begin{figure*}
\centering
\includegraphics[width=16cm]{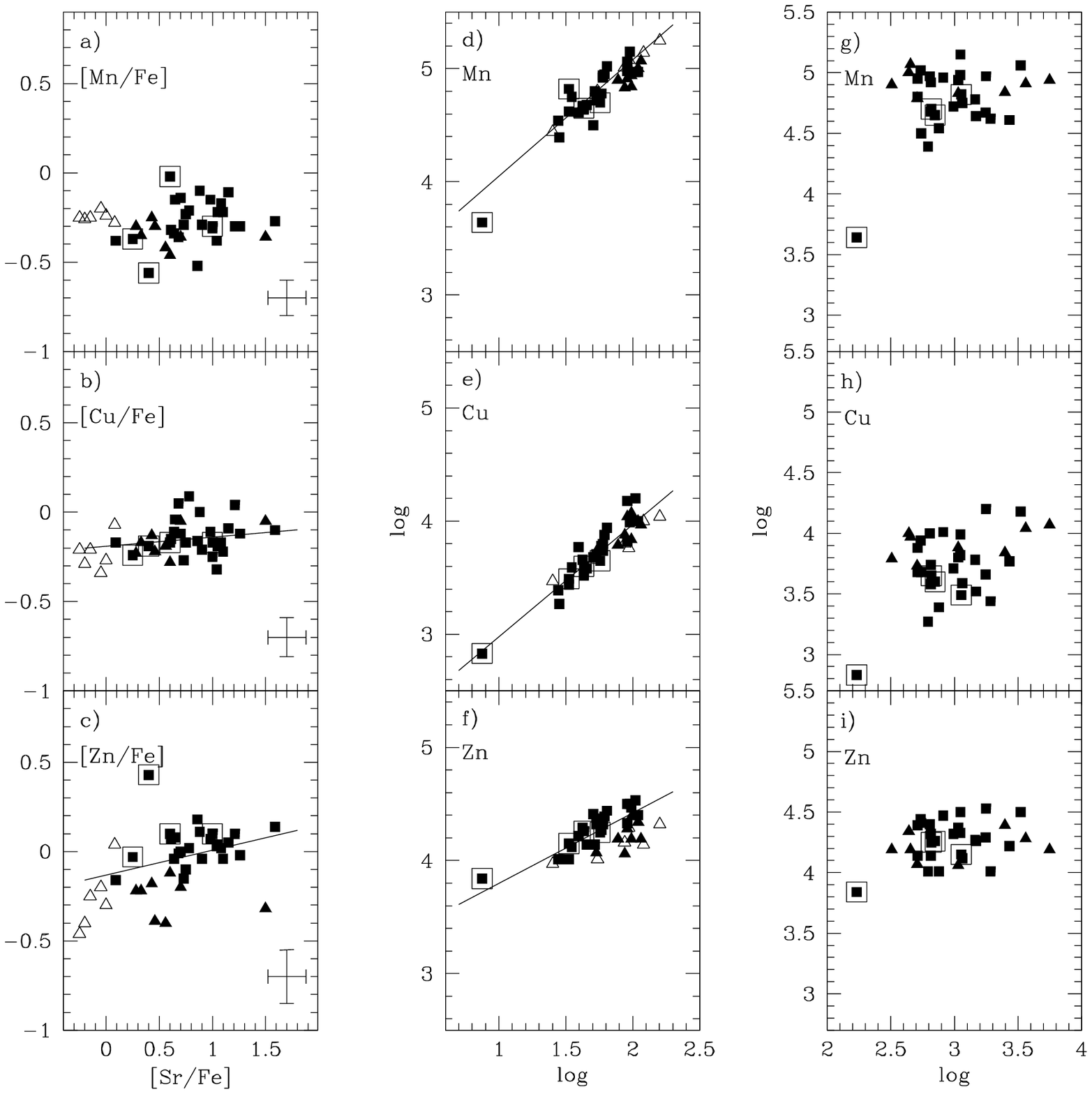}
\caption{Symbols are the same as in Fig. \ref{12356fg7}.
Panels a), b) and c) show [Mn,Cu,Zn/Fe] vs. [Sr/Fe]. The least-squares fits used only filled symbols.
Panels d), e) and f) show
$\log\epsilon$(X), where X is Mn, Cu, or Zn  as a function of the logarithimic part of the fraction
of Sr abundance with regard to
the weak component of the $s$-process. The least-squares fits used all symbols.
Panels g), h) and i) show  $\log\epsilon$(X), where X is Mn, Cu, or Zn as a function of the
logarithimic part of the fraction of Ba abundance relative to
the main component of the $s$-process. The least-squares fits used only filled symbols.
The results for the least-squares fits are given in Table \ref{fits}.}
\label{12356fg8}
\end{figure*}
}
%---------------------------------------------------------------------------------------------

\subsection{Production of iron peak elements in SN II}

Recently, \citet{feltzing07} obtained the run of [Mn/O] against
[O/H] in an attempt to better separate the contributions of SN Ia
and SN II to the synthesis of Mn, because oxygen is a virtually
exclusive product of SN II, while the Fe yield has
contributions from both types of SN \citep{timmes95}. This
interpretation is well confirmed by the behavior of oxygen when
related to europium in our sample stars, as shown in Fig.
\ref{12356fg6}: regarding [O/Eu] vs. [Fe/H], barium and normal stars
are mixed and the scatter is smaller than in the [O/Ba] vs. [Fe/H]
plot, where barium and normal stars are clearly segregated. As Eu
and O are believed to be released to the interstellar medium in
the same class of event, type II SNe, a flat [O/Eu] vs. [Fe/H] is
expected. Yet, considering barium, an $s$-element, there is no
clear correlation between [O/Ba] and [Fe/H]. Note that the normal
stars lie above all other data because their Ba abundances are much
lower when compared to the barium stars, while the abundance of O
is the same for both. Whereas the [Mn,Cu/O] vs. [O/H] runs in our
Fig. \ref{12356fg7} are essentially flat, [Zn/Fe] vs. [O/H] shows a
clearly decreasing trend.

Figure 13 of \citet{feltzing07} shows that the run of [Mn/O] vs.
[O/H] is flat up to [O/H] = $-$0.5 for halo and metal-poor
thick-disk stars, indicating that the yields of Mn and O are well
balanced in this interval. For [O/H] $\geq -$0.5, [Mn/O] vs. [O/H]
shows an increasing trend. According to \citet{bensby04}, the
archetypal signature of SN Ia in the thick disk does not occur up
to [O/H] $=$ 0, so this increasing trend could be attributed to
metallicity dependent Mn yields in SN II, the SN Ia contributing to
the synthesis of Mn only later. However, they also concede that
the contribution of SN Ia to the rise of Mn in the thin disk cannot
be completely neglected. The increasing trend of [Mn/O] vs. [O/H]
as shown in Fig. 13 of \citet{feltzing07} for [O/H] $\ge -$ 0.5 is
not seen in our Fig. \ref{12356fg7}, even considering that most of
our stars belong to the disk, according to \citet{menn97}.
Therefore, if one accepts that the nucleosynthetic disk
signature of SN Ia does not occur before [O/H] $=$ 0, the decrease
of [Zn/O] in our Fig. \ref{12356fg7} could be more directly
interpreted as caused by the action of metallicity-dependent yields
in massive stars between these two elements, and concurring with
our previous conclusion that a larger fraction of the synthesis of
Zn is owed to massive stars than is the case for Cu.

Concerning Mn, the previous absence of correlation between [Mn/Fe]
and the r-process elements in our Fig. \ref{12356fg5} and Fig.
\ref{12356fg4}, along with the flat [Mn/O] vs. [O/H] (Fig.
\ref{12356fg7}), again strengthens the interpretation that little is
owed to massive stars in the synthesis of Mn, again
attributing the bulk if the synthesis of oxygen and the r-process
elements to massive stars. Most of the Mn synthesis would then be
due to SN Ia, in line with the views of \citet{nissen00} and
\citet{carretta04}.

\subsection{Main and weak s-components}\label{weakmain}

Given that Mn, Cu, and Zn are thought to be partly produced by the
weak component of the $s$-process, and Sr also has an important
contribution from this component \citep{lugaro03}, we also
investigated how the abundances of all these elements are
related. At first we created [Mn,Cu,Zn/Fe] vs. [Sr/Fe] plots as
shown in panels a), b), and c) of Fig. \ref{12356fg8}. [Mn/Fe] vs.
[Sr/Fe] seems to have a flat trend, while [Cu/Fe] vs.
[Sr/Fe] yields an increasing trend, which is even stronger for
[Zn/Fe] vs. [Sr/Fe]. Next, we related the abundances of Mn, Cu, and
Zn ($\log\epsilon$) to the logarithm of the abundance fraction
correspondent to the weak component of the $s$-process for Sr, as
shown in panels d), e), and f) of Fig. \ref{12356fg8}. For the
stars of our work, the separation of the abundances
according to the nucleosynthetic process was made as in
\citet{di06b}, and the values will be shown in a forthcoming
paper. The values of $\chi^2_{red}$ shown in Table \ref{fits},
as well as the uncertaintities in the
correlation coefficients, tell us that there are increasing
and statistically significant correlations, confirming that
significant fractions of Mn, Cu, and Zn are produced by the weak
component of the $s$-process. In this case, the normal stars can
be included given that there is no difference in the separation of
components for both normal and barium stars. 

To investigate how the abundances of Mn, Cu, and Zn are related to
the abundance fraction corresponding to the main component of the
$s$-process \citep{di06b} we have added panels g), h), and
i) of Fig. \ref{12356fg8}. Normal stars were obviously
excluded from these panels. If Cu was preferentially depleted to
produce Ba through the main component of the $s$-process we would
see a decreasing trend, but the opposite is seen instead. 
That the [Cu/Fe] ratio is mostly flat when related to neutron
capture elements indicates that Cu is little, if at all,
affected by the $s$-process in the possible role of a seed
nucleus. On the other hand, considering the $s$-process as a chain
starting in $^{56}$Fe and ending in Bi, Cu is also in the
$s$-process path, so it is produced as well as destroyed, and our
results suggest that it is being preserved after all. A
very similar effect is seen for Zn. Our results agree
well with the assertion by \citet{matte93} that only a
very low contribution of the main component of the $s$-process
goes to the abundances of Zn and Cu, whereas the contribution of
the weak component is higher.

Lastly, it still stands out in the case of Cu that the two
aforementioned metal-poor, yellow symbiotic stars of the halo,
Hen 2-467 and BD-21$\degr$3873 analyzed by \citet{psc98} and
\citet{pp97}, respectively, do present a very low [Cu/Fe]
ratio that remains unexplained. Perhaps these objects have
something interesting to reveal about the operation of the
$s$-process operating in a regime different from most Ba-rich
stars discussed here.

%______________________________________________________________

\section{Conclusions}\label{concl}

We presented previously unpublished Mn abundances based on 
spectrum synthesis for the barium
star sample of \citet{di06a} and new C, N, O, Mn, Cu, Zn and heavy
element abundances for the barium star sample of \citet{rod07}.
Because barium stars show high overabundances in C and N, abundances
for these elements were also derived before those of the heavy
elements, because their lines are often affected by molecular bands
of CH and CN.

We were able to clearly establish in our analysis that the
stars considered to be normal by \citet{rod07} show very low abundances
of the $s$-elements when compared to bona fide barium
stars, while this difference is much reduced for the $r$-elements.
This confirms the previous analysis that these stars are normal
instead of barium stars, which further indicates that
additional work on proposed barium stars, that are classified
based on old data is necessary in order to check whether 
they indeed belong to this class.

The [Mn,Cu,Zn/Fe] abundance ratios were correlated to elements that 
are nucleosynthetically dominated by the $s$-process (Ba, Y, Nd) and
the $r$-process (Eu, Gd, Dy). The runs of [Mn/Fe] vs.
[$s$-,r-elements/Fe] show much weaker correlations when compared
to those of [Cu,Zn/Fe] vs. [$s$-,r-elements/Fe]. The absent
correlation of [Mn/Fe] with the $r$- and $s$-element abundances
along with the flat [Mn/O] vs. [O/H] indicate that the production
of Mn is a mix of several processes, likely to be dominated
by SN Ia nucleosynthesis, which agrees well with other
works in the literature. The weak correlation between Cu and the
$r$-elements indicates that the bulk of Cu production does not
come from SN II. The increasing trend of [Zn/Fe] vs. [Eu/Fe]
indicates that for Zn a higher abundance fraction is owed to
massive stars than in the case of Cu.

Our results suggest that significant fractions of Mn, Cu, and Zn
are produced by the weak component of the $s$-process based
on the increasing correlations found for Mn, Cu, and Zn abundances
and the abundance fraction due to this process to the Sr
abundance. On the other hand, the near absence of correlation
between Mn, Cu, and Zn abundances and the abundance fraction of Ba
due to the main component of the $s$-process suggests that only a
very low contribution of this component goes to the abundances
of Mn, Zn, and Cu. We found in particular that relations
between the [Cu/Fe] ratio and those owed to neutron capture
elements are mostly flat, suggesting that Cu is little if at all
affected by the $s$-process in the possible role of a seed
nucleus. This could indicate that Mn, Cu, and Zn are essentially
preserved during the main $s$-processing in AGB stars. Our results
point toward a significant contribution of the weak
$s$-processing to the nucleosynthesis of Mn, Cu, and Zn.

Nevertheless, the complex behavior of the abundances of
these elements remain unaccounted for by theoretical methods in
its details. More abundance data, both for normal and chemically
peculiar objects, would certainly contribute toward a better
understanding of the nucleosynthesis of the elements in the
transition region between the Fe-peak and neutron capture
species.

%
%______________________________________________________________

\begin{acknowledgements}

DMA acknowledges the following post-doctoral fellowships:
FAPERJ n$^{\circ}$ 152.680/2004, CAPES n$^{\circ}$ BEX 3448/06-1,
and FAPESP n$^{\circ}$ 2008/01265-0. The STRI
at the University of Hertfordshire, where part of this work was
completed, is acknoledged by DMA. GFPM acknowledges financial support by
CNPq grant n$^{\circ}$ 476909/2006-6, FAPERJ grant n$^{\circ}$
APQ1/26/170.687/2004, and a CAPES post-doctoral fellowship
n$^{\circ}$ BEX 4261/07-0. We are grateful to Licio da Silva and
Beatriz Barbuy for making available some of the spectra, and to
Beatriz Barbuy for making available the spectrum synthesis code.
We are also grateful for valuable suggestions from an anonymous referee
and to Marcelo Porto Allen for his useful comments.
Use was made of the Simbad database, operated at CDS, Strasbourg,
France, and of NASA's Astrophysics Data System Bibliographic
Services.

\end{acknowledgements}
%
%______________________________________________________________

\end{document}